\documentclass[aps,prX,twocolumn,amsmath,amssymb,longbibliography]{revtex4-1}

\usepackage{graphicx}
\usepackage{dcolumn}
\usepackage{bm}
\usepackage{multirow}
\usepackage[utf8]{inputenc}

\usepackage{color}

\usepackage[dvipsnames]{xcolor}

\begin{document}

\title{
Orbitally Dominated Rashba-Edelstein Effect in Noncentrosymmetric Antiferromagnets} 

\author{Leandro Salemi}
 \email{leandro.salemi@physics.uu.se}
\author{Marco Berritta}%
\author{Ashis K.\ Nandy}
\altaffiliation{Current address: School of Physical Sciences, National Institute of Science Education and Research, HBNI, Jatni 752050, India.}
\author{Peter M.\ Oppeneer}%
\affiliation{%
 Department of Physics and Astronomy, Uppsala University, P.\,O.\ Box 516, S-751 20 Uppsala, Sweden}%

\date{\today}

\begin{abstract}
Efficient manipulation of magnetic order with electric current pulses is desirable for achieving fast spintronic devices. The Rashba-Edelstein effect, 
wherein a
spin polarization is electrically induced in noncentrosymmetric systems, 
provides a 
mean to achieve current-induced staggered spin-orbit torques.
Initially predicted for spin, the orbital counterpart of this effect has been disregarded up to now. 
Here, we present a generalized Rashba-Edelstein effect, which generates not only spin polarization but also orbital polarization, which we find to be 
far from being negligible and could play a crucial role in the magnetization dynamics. We 
show that the orbital Rashba-Edelstein effect does not require spin-orbit coupling to exist. 
We present first-principles calculations of the frequency-dependent spin and orbital Rashba-Edelstein susceptibility tensors for the noncentrosymmetric antiferromagnets CuMnAs and Mn$_2$Au. 
We show that the electrically induced local magnetization has both staggered in-plane components and non-staggered out-of-plane components, and can exhibit Rashba-like or Dresselhaus-like symmetries, depending on the magnetic configuration.
Furthermore, there is an induced local magnetization on the nonmagnetic atoms as well, that is smaller in Mn$_2$Au than in CuMnAs. We compute sizable induced magnetizations at optical frequencies, which suggest that electric-field {driven} switching could be achieved at much higher frequencies.
\end{abstract}

\maketitle

\section{Introduction}
The efficient manipulation of the magnetization of materials remains a crucial challenge in the field of spintronics \cite{Zutic2004,Brataas2012,Hellman2017}.
Although previously disregarded, antiferromagnets have recently emerged as appealing candidate materials for  information storage devices since they offer various advantages \cite{Jungwirth2016,Nemec2018,Baltz2018}. Specifically, antiferromagnets are robust against external magnetic field perturbations, they are available as insulators, semiconductors and metals, allowing thus for versatile environment integration, and they often have a high N{\'e}el temperature \cite{Maca2012,Yamaoka1974,Barthem2013}, suitable for room-temperature operation of the devices. Moreover, their intrinsic spin dynamics is ultrafast, in the THz domain \cite{Kimel2004,Fiebig2008,Kampfrath2010}
(compared to GHz dynamics reported for ferromagnets \cite{Pashkin2013,Kirilyuk2010}).

Achieving efficient control over the magnetization in antiferromagnets is however an entirely different issue.
The spin Hall effect has proven to generate a spin-polarized current \cite{Kato2004a,Sinova2015} and thereby create a spin-orbit torque that can act efficiently on the magnetization of a ferromagnetic layer \cite{Miron2011,Liu2012,Garello2013,Baumgartner2017}. A different effect, the Rashba-Edelstein effect (REE) has been proposed as a method to induce a nonequilibrium spin polarization through an electrical current in solids lacking inversion symmetry \cite{Edelstein1990}. 
This effect (also called inverse spin galvanic effect) was initially predicted in 1990 by Edelstein \cite{Edelstein1990} using a Rashba spin-orbit coupling (SOC) \cite{Bychkov1984}; it was experimentally observed in GaAs heterostructures \cite{Kato2004,Silov2004,Ganichev2006}. 

More recently, the Rashba-Edelstein effect has been proposed as a method to create a current-induced staggered spin polarization and spin-orbit torque in the noncentrosymmetric antiferromagnets CuMnAs and Mn$_2$Au, causing the antiferromagnetic magnetic moments to flip to a {perpendicular} direction \cite{Wadley2016,Olejnik2017,Bodnar2018,Meinert2018,Godinho2018}. These recent experiments  have shown that current driven switching of the N{\'e}el vector is possible, however, the operation of the spin-orbit torque and the switching path are not understood yet. Microscopic investigations indicate that a complex switching process with both domain wall motion and domain flips may occur
 \cite{Grzybowski2017,Wadley2018}. It is moreover a question how large the induced staggered moments are. So far, linear-response tight-binding calculations with Rashba SOC \cite{Zelezny2014,Zelezny2017} and an \textit{ab initio} calculation \cite{Wadley2016} of the current induced magnetic fields have been performed, that however differed considerably. Also a semiclassical model based on the Boltzmann equation has been employed to compute the induced magnetization in a Weyl semimetal \cite{Johansson2018}. These investigations concentrate moreover on the induced spin polarization and neglect any possible contribution stemming from an induced orbital magnetization.

In the work, we investigate computationally the full magnetic polarization induced by an applied electric field in the noncentrosymmetric antiferromagnets CuMnAs and Mn$_2$Au. To this end we employ an accurate full-potential, all-electron code WIEN2k \cite{Blaha2018} to compute both the spin REE (SREE) and orbital REE (OREE) tensors within the density-functional theory (DFT) framework. These tensors are computed over a wide frequency spectrum, i.e.\ we do not restrict the calculation to the case of static electric fields.
Our calculations bring new insights into the Rashba-Edelstein effect in these antiferromagnets. We find that the dominant contribution to the induced polarizations stems from the orbital REE. The OREE tensor can have a symmetry \textit{different} from that of the SREE tensor (e.g., Rashba-type vs.\ Dresselhaus-type of symmetry). Due to the pronounced Rashba symmetry of the OREE tensor, a strong orthogonal orbital-momentum locking is obtained for in-plane electric fields.
We find furthermore that quite sizable moments can be {electrically} induced on the nonmagnetic atoms. Investigating the origin of the large induced orbital polarizations, we show that these are present even without spin-orbit interaction, whereas the spin REE tensor is proportional to the SOC and vanishes without SOC, signifying that the latter are induced  through the relativistic SOC, whereas the former have a nonrelativistic origin.

\section{Theoretical Framework} 
We use linear-response theory to evaluate the magnetic response to a time-dependent electric field $\bm{E}(t)$.
The  induced magnetic polarization $\delta \bm{M} = \mu_B \delta ( \bm{L} + 2  \bm{S})$ in the frequency domain 
reads
\begin{equation}
\delta M_i(\omega) = \sum_j  \Big( \chi^L_{ij}(\omega) + 2 \chi^S_{ij}(\omega) \Big) E_j(\omega) ,
\label{ind-mag}
\end{equation}
where $\chi^L_{ij}(\omega)$ and $\chi^S_{ij}(\omega)$ ($i = x,\, y,\, z$) are the orbital and spin Rashba-Edelstein susceptibility tensor, respectively (in units of $ \mu_B$\,nm/V).
These tensors can be derived by considering  the influence of the time-varying electric field as a perturbation $\hat{V}(t) = -e \bm{E}(t) \cdot \bm{\hat{r}}$ within a periodic crystal potential, with $e$ the electron charge and $\bm{\hat{r}}$ the position operator, to the unperturbed time-independent Kohn-Sham Hamiltonian  $\hat{H}_0$. 
Employing linear-response theory the tensors can be expressed in terms of the solutions of the unperturbed Kohn-Sham Hamiltonian as
\begin{eqnarray}
\label{eq:Chi}
\chi_{_{\alpha\beta}}^{B}(\omega) &=&  \frac{i e}{ m_e} \int_{\Omega} \frac{d\bm{k}}{\Omega} \sum_{n \neq m}  \frac{ f_{m\bm{k}} - f_{n\bm{k}} }{ \hbar \omega_{nm\bm{k}}} 
\frac{ B^\alpha_{mn\bm{k}} ~ p^\beta_{nm\bm{k}} }{ \omega - \omega_{nm\bm{k}} + i \tau_{\rm inter}^{-1}}  \nonumber \\
&-& \frac{ie}{m_e} \int_\Omega \frac{d \bm{k}}{\Omega} ~ \sum_{n} 
\frac{\partial f_{n\bm{k}}}{\partial \epsilon} \frac{B^\alpha_{nn \bm{k}}~~p^\beta_{nn \bm{k}} }{\omega + i \tau_{\text{intra}}^{-1}} ,
\end{eqnarray}
where $f_{n\bm{k}}$ is the occupation of Kohn-Sham state $|n\bm{k}\rangle$ with energy $\epsilon_{n\bm{k}}$ at wavevector $\bm{k}$, $\Omega$ the Brillouin zone volume, $\bm{p}_{nm\bm{k}}$  are the momentum-operator matrix elements and $\bm{B}_{nm\bm{k}}$ are the matrix elements of the spin ($\hat{\bm{S}}$) or orbital angular momentum ($\hat{\bm{L}}$) operator, respectively, and
$ \hbar \omega_{nm\bm{k}} = \epsilon_{n\bm{k}} - \epsilon_{m\bm{k}}$.  

The REE tensor (\ref{eq:Chi}) contains two distinct contributions, interband ($n \neq m$) and intraband ($n=m$) contributions. The former describe transitions between the valence and conductions states, the latter describes the response of electrons around the Fermi energy. 
$\tau^{-1}$ is a broadening parameter that accounts for the finite electron-state lifetime; it can be different for intraband and interband transitions.  
Here, for sake of simplicity, we use the same value of $\tau$ for the interband and intraband contributions (see Appendix), because our aim is to understand the role and symmetry of the spin and orbital Rashba-Edelstein effect and not their precise value.
%
We {further} note that our interband formulation is different from another recent investigation  \cite{Zelezny2014,Zelezny2017},
in which  the expression
$\frac{ f_{m\bm{k}} - f_{n\bm{k}}} {\omega_{nm\bm{k}}  +i/ \tau}$ is used,
but the analytic dependence in Eq.\ (\ref{eq:Chi}) guarantees that the response is causal and that Kramers-Kronig relations between real and imaginary part are fulfilled, which is necessary for {the REE at} nonzero frequencies.

\begin{figure*}[ht]
\includegraphics[width=\linewidth]{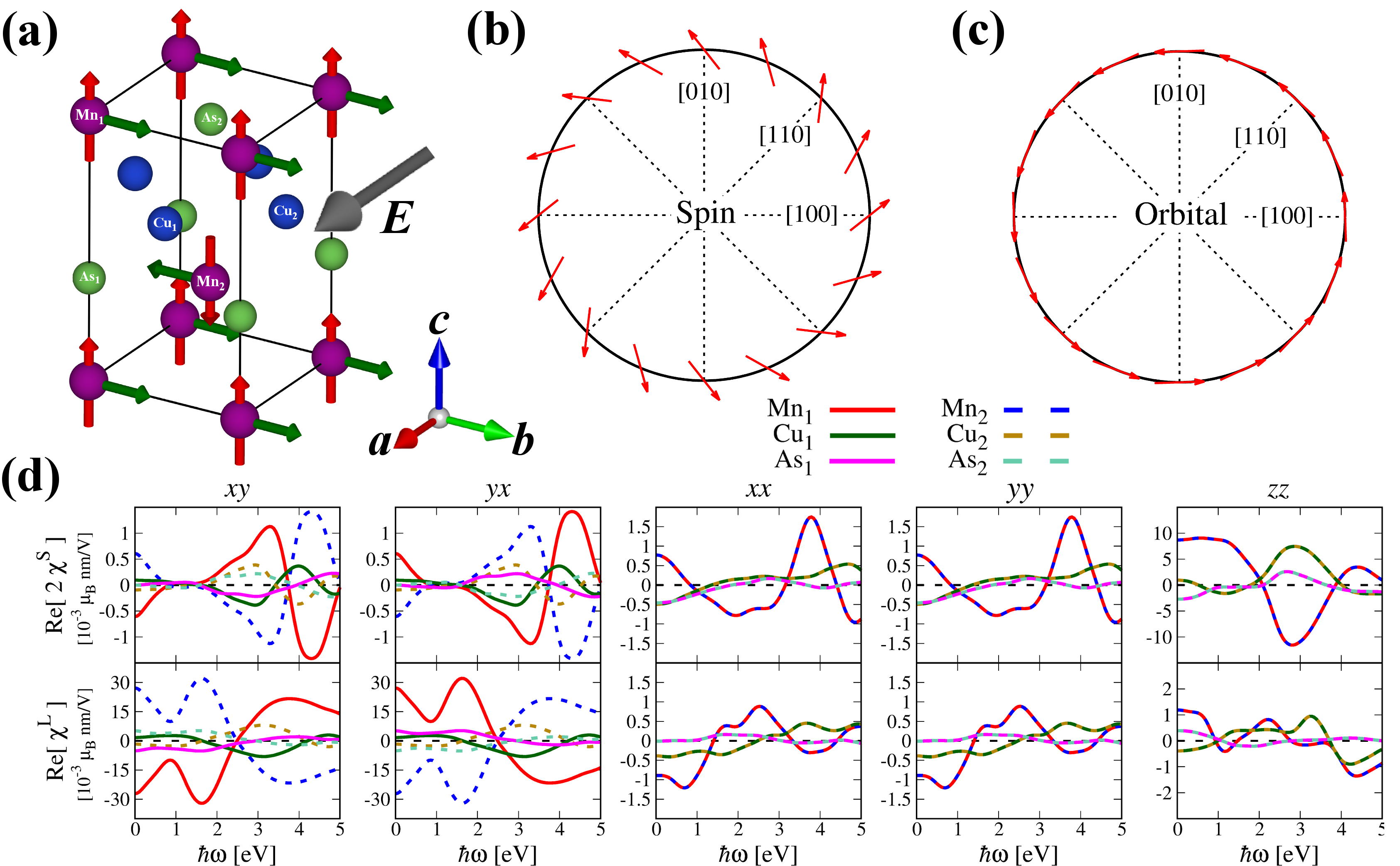}
\caption{{Magnetization induced by the Rashba-Edelstein effect in antiferromagnetic CuMnAs.}
{(a)} Sketch of the tetragonal unit cell of CuMnAs 
with the magnetic moments constrained along the $c$-axis. 
The red arrows on the manganese atoms represent the initial magnetic {moments}. Applying an electric field $E$ along the (100) direction (grey arrow) induces a non-equilibrium magnetization mainly along the (010) direction (green arrows). {(b)} Symmetry of the induced spin magnetization as a function of the static electric field direction for Mn$_1$. {(c)} Symmetry of the induced orbital magnetization as a function of the static electric field direction for Mn$_1$. {(d)} Real part of the frequency-dependent spin and orbital Rashba-Edelstein susceptibility. Only the real part of the nonzero tensor components is displayed.}
\label{fig:Fig1}
\end{figure*}

\begin{figure*}[ht]
\centering
\includegraphics[width=\linewidth]{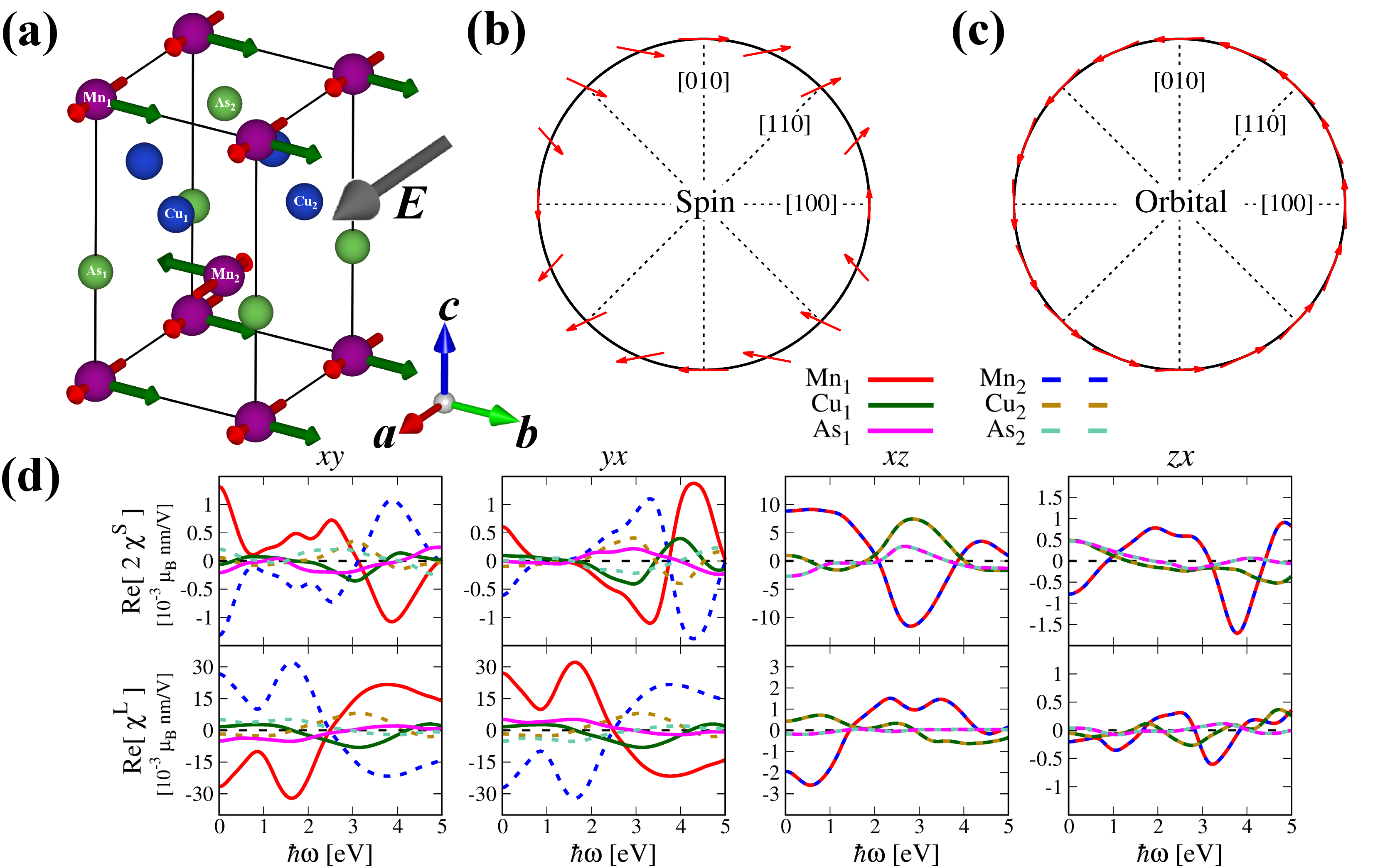}
\caption{ Rashba-Edelstein effect in CuMnAs with magnetic moments along the $a$-axis.
{(a)} Sketch of the tetragonal unit cell of CuMnAs.  The red arrows on the Mn atoms represent the initial magnetic {moments}.
Applying an electric field $E$ along the (100) direction (grey arrow) induces a nonequilibrium magnetization mainly along the (010) direction (green arrows). {(b)} Symmetry of the induced spin magnetization as a function of the static electric field direction for Mn$_1$. {(c)} Symmetry of the induced orbital magnetization as a function of the static electric field direction for Mn$_1$. {(d)} Real part of the frequency-dependent spin and orbital Rashba-Edelstein susceptibility tensors. Only the real part of the nonzero components is displayed.}
\label{fig:Fig2}
\end{figure*}

\begin{figure*}[ht]
\centering
\includegraphics[width=\linewidth]{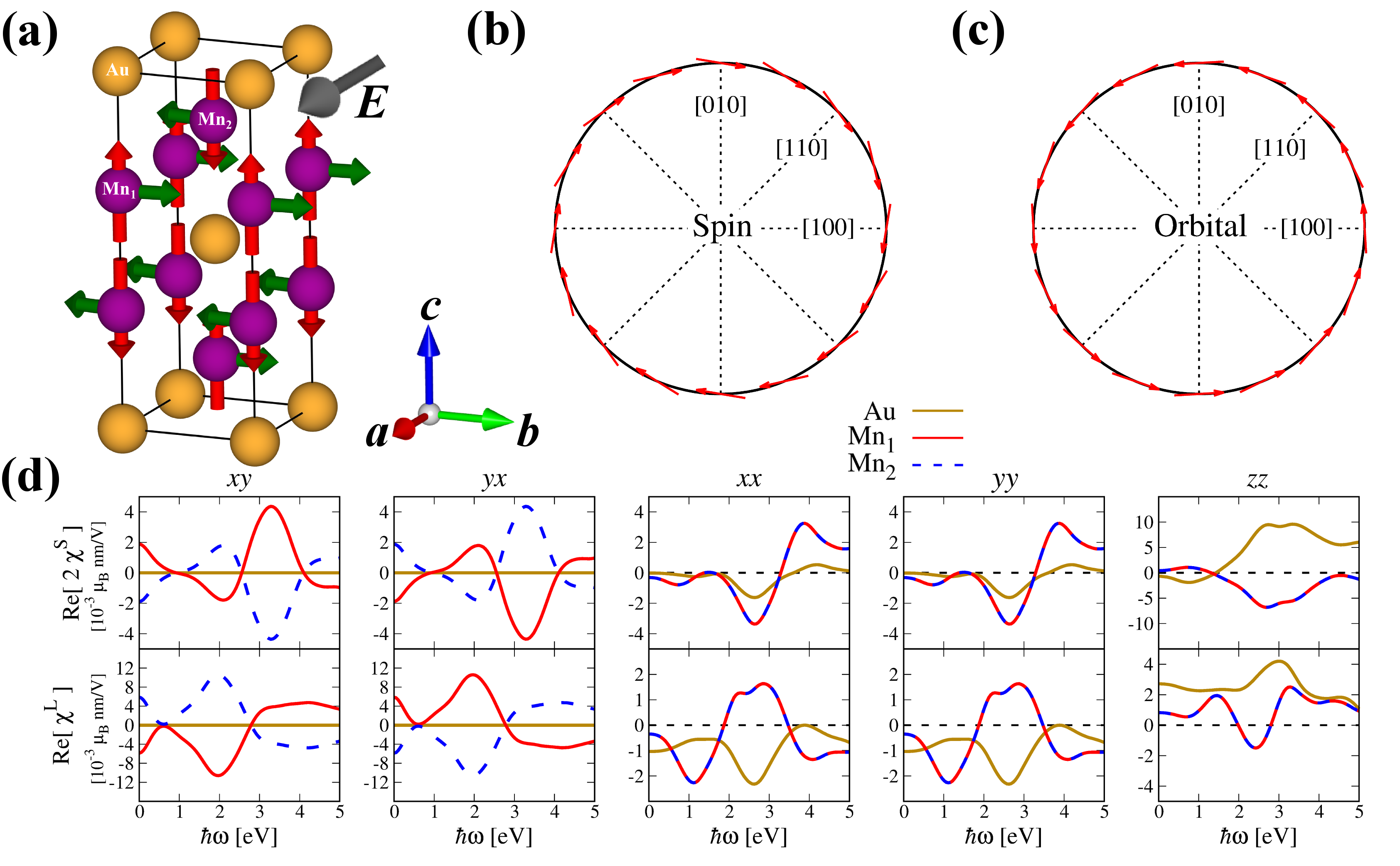}
\caption{Theory of Rashba-Edelstein effect in Mn$_2$Au with magnetic moments along the $c$-axis.
{(a)} The unit cell of Mn$_2$Au, with red arrows on the Mn atoms depicting the initial magnetic {moments}. Applying an electric field $E$ along the (100) direction (grey arrow) induces a non-equilibrium magnetization tilted in between the (010) and (100) direction (green arrows). {(b)} Symmetry of the induced spin magnetization as a function of the static electric field direction for Mn$_1$. {(c)} Symmetry of the induced orbital magnetization as a function of the static electric field direction for Mn$_1$. {(d)} Real part of the frequency-dependent spin and orbital Rashba-Edelstein susceptibility. Only the real part of the nonzero components is displayed.}
\label{fig:Fig4}
\end{figure*}

\begin{figure*}[ht]
\centering
\includegraphics[width=\linewidth]{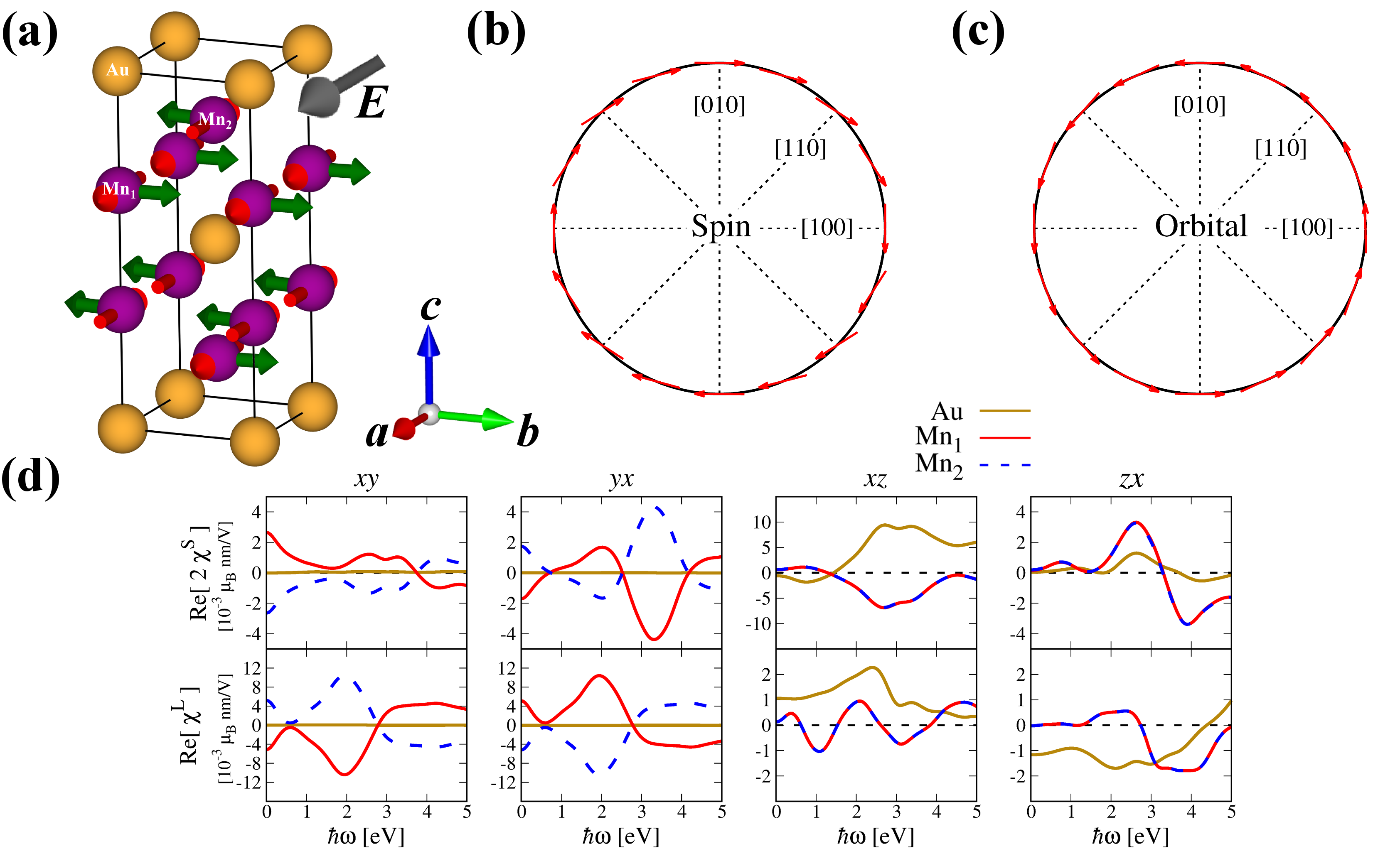}
\caption{Rashba-Edelstein effect in Mn$_2$Au with magnetic moments along the $a$-axis.
{(a)} The unit cell of Mn$_2$Au.  The red arrows on the Mn atoms represent the initial magnetic {moments}. Applying an electric field $E$ along the (100) direction (grey arrow) induces a nonequilibrium magnetization mainly along the (010) direction (green arrows). {(b)} Symmetry of the induced spin magnetization as a function of the static electric field direction for Mn$_1$. {(c)} Symmetry of the induced orbital magnetization as a function of the static electric field direction for Mn$_1$. {(d)} Real part of the frequency-dependent spin and orbital Rashba-Edelstein susceptibility. Only the real part of the nonzero components is displayed.}
\label{fig:Fig5}
\end{figure*}

\section{Results}

To evaluate the
frequency-dependent SREE and OREE tensors, we adopt the DFT formalism as implemented in the full-potential linearized augmented plane-wave (FLAPW) all-electron code WIEN2k \cite{Blaha2018}. Details of the computational approach are given in the Appendix. 
In the following we apply this framework to  noncentrosymmetric CuMnAs and Mn$_2$Au that have recently drawn attention for antiferromagnetic spintronics \cite{Jungwirth2016,Wadley2016,Olejnik2018,Bodnar2018,Meinert2018,Baltz2018}.

\subsection{CuMnAs}

Our DFT calculations give that CuMnAs has an antiferromagnetic ground state with Mn atoms carrying a magnetic moment of $\sim$$3.66~\mu_B$,  in agreement with recent experiments \cite{Wadley2013,Wadley2015}.  The tetragonal cell of CuMnAs (space group $P4/nmm$), shown in Fig.\ \ref{fig:Fig1}{(a)}, consists of six inequivalent atoms, two of each chemical species. {Both the Mn and As atoms have  the $4mm$ point group whereas the Cu atoms possess the $-4m2$ point group symmetry.} The magnetic ordering is such that adjacent \{001\}  Mn planes are antiferromagnetically coupled while Mn atoms laying in the same plane are ferromagnetically ordered. The As atoms are also found to carry a small magnetic moment of $\sim$$2.33~10^{-3}~\mu_B$. Their orientation is such that \{001\} As planes are ferromagnetically coupled to the closest \{001\} Mn plane. The Cu atoms are found to be non-magnetic. 

The AFM moments can orient along different N{\'e}el vector axes and this direction of the AFM moments depends sensitively, for thin films, on the interplay of intrinsic magneto-crystalline anisotropy and shape anisotropy. 
Experimentally, an orientation of the spins in the $ab$-plane has been observed for thin films \cite{Wadley2015}. 

The REE tensors depend on the orientation of the moments. We compute them here for different orientations of the moments, and in addition,  we compute the atom-resolved tensors' spectra, using a 
specific labeling of the atoms as shown in Fig.\ \ref{fig:Fig1}{(a)}.

We first consider the case where magnetic moments are oriented along the $c$-axis and the applied field is along the $a$-axis
 (Fig.\ \ref{fig:Fig1}{(a)}).
This magnetic configuration does not break the 4-fold rotational symmetry about the $c$-axis (hard magnetization axis). The nonzero components of the atom-resolved spin and orbital Rashba-Edelstein tensors are displayed in Fig.\ \ref{fig:Fig1}{(d)}. 
Several remarkable observations can now be made. First, there are frequency-dependent induced moments not only on the Mn atoms, but also on the Cu and As atoms. Second, the orbital contribution that was thus far disregarded, is not negligible. In fact the staggered orbital part $\chi_{xy}^{L}$ is the dominating part of the response and is $\sim$60 times larger than its spin counter part at $\omega = 0$. {In the near-infrared region ($\hbar \omega =0.9$ to 1.7 eV) $\chi_{xy}^L$ dominates even more, since the spin response $\chi_{xy}^S$ is almost zero.} Third, apart from the \emph{staggered} components, that are such that antiferromagnetic Mn1 and Mn2 atoms experience an opposite response (off-diagonal $xy$ and $yx$), there are also \emph{homogeneous} induced components that act in the same direction for a given atomic species (see diagonal diagonal $xx$, $yy$ and $zz$ tensors elements in Fig.\ \ref{fig:Fig1}{(d)}). These \textit{non-staggered} induced magnetizations can alter the atomic torques and influence eventual spin switching.
{Lastly, we note that SREE and OREE tensors of the individual elements obey different symmetries, specific to the atomic site's point group.}
For the Mn atoms we observe $\chi^{S,L}_{xy} = -\chi^{S,L}_{yx}$, and $\chi^{S,L}_{xx} = \chi^{S,L}_{yy}$.
 The same symmetry of the tensors is obtained for the As atoms, but for the Cu atoms $\chi^{S,L}_{xy} = \chi^{S,L}_{yx}$ and $\chi^{S,L}_{xx} = \chi^{S,L}_{yy}$.

The calculated orientation of the induced moments as a function of the direction of an in-plane applied \textit{static} electric field is displayed in Figs.\ \ref{fig:Fig1}{(b)} and \ref{fig:Fig1}{(c)} for the spin and orbital part, respectively. 
We observe a nearly Rashba-like behavior for the spin part with nonorthogonal spin-momentum locking, whereas the orbital part possesses {a perfect Rashba symmetry} with orthogonal orbital-momentum locking
(for a definition, see e.g.\ \cite{Manchon2015,Ciccarelli2016}). These plots are obtained by computing the tensors at $\omega =0$ while 
varying the direction of $\bm{E}$. It is important to note that the induced spin and orbital moments depend on the frequency $\omega$.
In addition, the fact that the  spin and orbital polarization are induced in different directions, {and can even be antiparallel (see below)} has {an} important consequence. {The resultant torque field that acts on the atomic moments in a Landau-Lifshitz-Gilbert spin-dynamics formulation}  can {then} not be represented in the form of a single atomic Zeeman {field, corresponding to} an interaction $(\mu_B / \hbar ) (\hat{\bm{L}} + 2\hat{\bm{S}})\cdot \bm{H}$, with $\bm{H}$ the applied {atomic} Zeeman magnetic field, as this would lead to a proportional induced spin and orbital atomic moment.

We now consider the case of CuMnAs with an in-plane magnetization along the (100) direction which corresponds to the magnetic structure realized in experiments. As shown in Fig.\ \ref{fig:Fig2}{(a)}, applying a static electric field (grey arrow) along the magnetization direction (red arrows) induces magnetic moments (green arrows) mainly on the Mn atoms. Those magnetic moments are mainly staggered, i.e.,\ they are practically antiparallel to each other for AFM coupled Mn atoms. However, a small parallel out-of-plane contribution is also present. This non-staggered feature of the magnetic response can be recognized by looking at the SREE and OREE tensors, shown in Fig.\ \ref{fig:Fig2}{(d)} (the nonzero $\chi_{zx}$ tensor components). Notably, the by far dominant part of the induced magnetic polarization is again contained in the staggered  $xy$ and $yx$ components of the orbital response.

Another important point to be noticed is that the nonzero homogeneous tensor components have changed with the changed direction of the N{\'e}el vector.
The non-staggered components for CuMnAs with magnetization along (001) were the diagonal $xx$, $yy$ and $zz$ components while for the magnetization along (100), these are the $xz$ and $zx$ components. As can be seen in Figs.\ \ref{fig:Fig1}(d) and \ref{fig:Fig2}(d), the computed SREE spectra are very similar, with an inverted sign {($\chi_{zz}^S \longrightarrow \chi_{xz}^S$, and $\chi_{xx}^S \longrightarrow -\chi_{zx}^S$).} This is a direct demonstration that the electrically induced magnetization depends on the underlying magnetization direction itself. This can be understood as an  influence of the magnetization direction on the eigenstates which affects the induced magnetization \cite{Zelezny2017}. This effect has also been observed experimentally in (Ga,Mn)As \cite{Kurebayashi2014}.  
Computing the symmetry of the momentum-dependent induced spin and orbital polarizations for an in-plane electric field, we find that the spin-resolved part exhibits a Dresselhaus-like symmetry whereas the orbital-resolved part exhibits Rashba symmetry
(Figs.\ \ref{fig:Fig2}{(b)} and \ref{fig:Fig2}{(c)}). Here, {it can be recognized that the} induced spin and orbital polarizations cooperate {for a static in-plane electric field along the $a$ axis ($[100]$)} and exert thus spin and orbital torques in the same direction. {However, for an in-plane field along the $b$-axis ($[010]$) the induced spin and orbital polarizations are opposite and thus act against each other.}
We further note that the symmetries of the REE tensor are now such that 
$\chi^S_{xy} \neq -\chi^S_{yx}$, but $\chi^L_{xy} = -\chi^L_{yx}$ for the Mn and As atoms. The latter tensor elements are the largest, which illustrates the dominance of the orbital REE.

\subsection{Mn$_2$Au}

Mn$_2$Au crystallizes in the tetragonal structure shown in Fig.\ \ref{fig:Fig4}{(a)} ($I4/mmm$ space group). The ground state of Mn$_2$Au is computed to be antiferromagnetic with magnetic moments of $3.69~\mu_B$ only on the manganese atoms. 
Experimentally, the magnetization of Mn$_2$Au films is found to lay in $\{001\}$ (basal) planes, with $\sim$4 $\mu_B$ moments on Mn \cite{Barthem2013}. The unit cell consists of two equivalent Au atoms and two pairs of inequivalent Mn atoms, labeled Mn$_1$ and Mn$_2$ in Fig.\ \ref{fig:Fig4}(a). {The four Mn atoms have the $4mm$ (polar) point group symmetry and the two Au atoms 
have 
the $4/mmm$ (centrosymmetric) point group symmetry.}

Figure \ref{fig:Fig4}{(d)} shows the nonzero SREE and OREE tensor elements, computed for Mn moments  along the $c$ axis. The calculated tensors exemplify that the REE of Mn$_2$Au is {in several aspects} different from that in CuMnAs.
In contrast to CuMnAs, the spin and orbital responses for both the $xy$ and $yx$ components are staggered in Mn$_2$Au and the homogeneous part of the response is in  the diagonal $xx$, $yy$ and $zz$ components, similar to CuMnAs. 
Also, while the orbital part of the response is far from being negligible, however, it is not as dominant as in the case of CuMnAs. The largest orbital in the off-diagonal elements is almost 3 times larger than the spin contribution for $\omega =0$.
We can furthermore observe that the non-magnetic Au atoms do not display any finite \textit{staggered} response, {consistent with the centrosymmetric nature of its point group $4/mmm$.}
%

The directional dependence of the current-induced moments on Mn atoms as a function of the direction of an in-plane applied static electric field is shown in Figs.\ \ref{fig:Fig4}{(b)} and \ref{fig:Fig4}{(c)} for the spin and orbital response, respectively. 
The spin response exhibits a
Rashba-like behavior and the orbital counterpart possesses {a Rashba symmetry, too, but notably opposite to that of the spin response.}
Hence, for any applied in-plane field, the current-induced spin and orbital moments will exert torques in {antiparallel} directions during a switching process.

We now consider Mn$_2$Au with moments laying in the $ab$-plane along the $(001)$ direction, see Fig.\ \ref{fig:Fig5}{(a)}. As for CuMnAs, the magnetic moments have been experimentally found to lay in the $ab$-plane \cite{Barthem2013}. Here, the calculated nonzero REE tensor elements, shown in Fig.\ \ref{fig:Fig5}{(d)}, are the $xy$, $yx$, $xz$ and $zx$ components. In this configuration, the staggered responses for both spin and orbital contributions {are} present in the $xy$ and $yx$ components whereas the non-staggered responses {are} present in the $zx$ and $xz$ components, that however give smaller contributions. 
The mainly staggered response corroborates the investigation of \v{Z}elzen{\'y} \textit{et al}.\ \cite{Zelezny2014}, who predicted staggered spin-orbit fields on the two Mn sublattices.
{The symmetries of the main staggered tensor elements are $\chi^S_{xy} \neq  -\chi^S_{yx}$ but $\chi^L_{xy} = -\chi^L_{yx}$, as we also obtained for CuMnAs with N{\'e}el vector along the $a$-axis.}

Considering the symmetry of the induced polarizations for an in-plane field in Figs.\ \ref{fig:Fig5}(b) and \ref{fig:Fig5}(c), we find that the SREE exhibits a {Rashba-type} behavior and the OREE exhibits a {pure Rashba} symmetry. {Again}, the possible non-cooperativity of the OREE and SREE when exerting a torque can be fully recognized. When both the static moments and electric field are along the $a$-axis {($[100]$)} the induced spin and orbital magnetizations are antiparallel and {the torques will} partially compensate each other. For an in-plane electric field $E$ along the $b$-axis 
the orbital and spin magnetizations do {also} counteract each other (see also Fig.\ \ref{fig:Fig5}{(d)}), but this {configuration} only leads to an induced moment along the static AFM moments that does not exert a local torque on the atomic moment. This exemplifies that devising optimal switching conditions can be quite intricate.


\begin{figure}[ht]
\centering
\includegraphics[width=\linewidth]{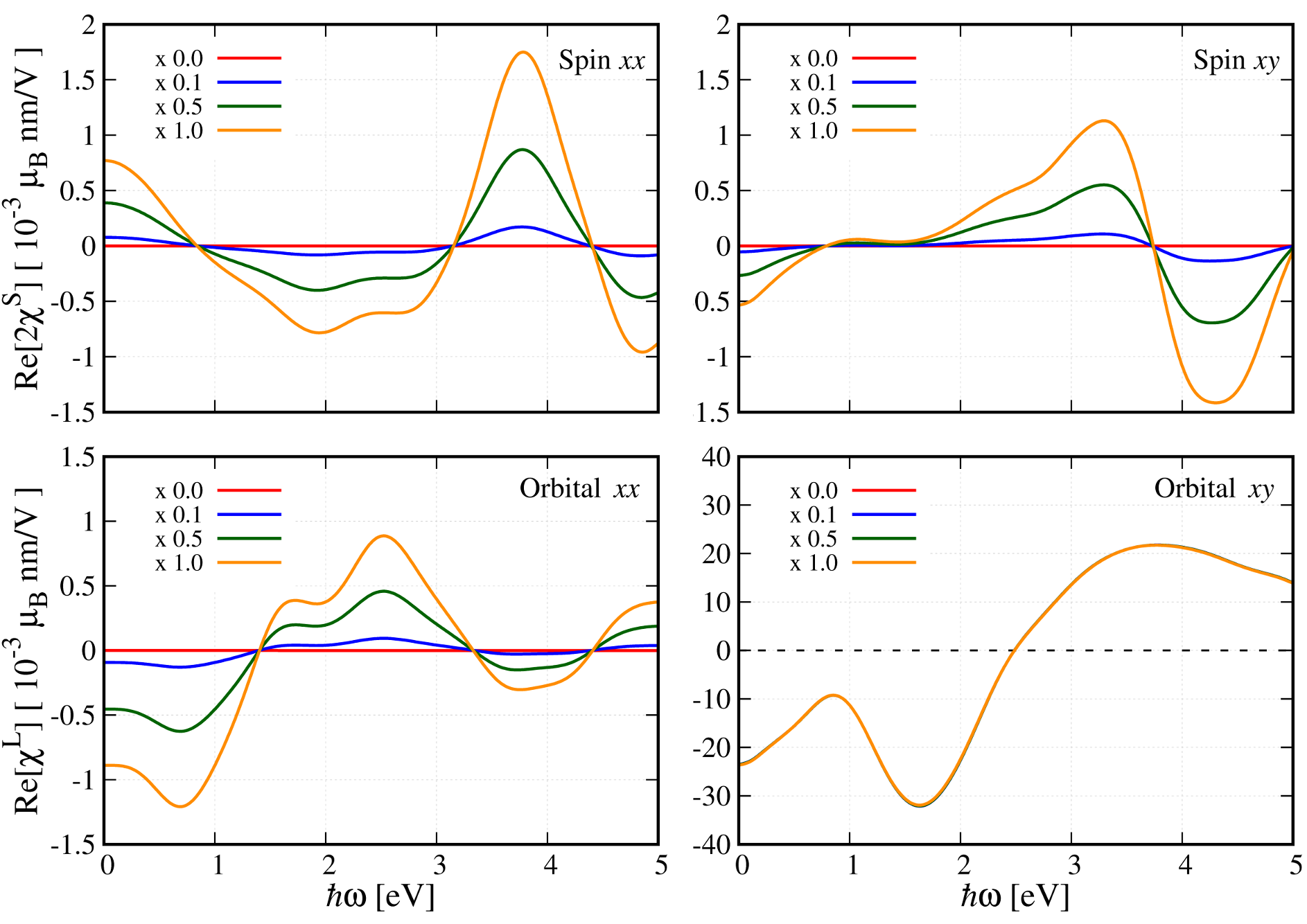}
\caption{{Rashba-Edelstein effect as a function of spin-orbit coupling strength computed for Mn1 atom of antiferromagnetic CuMnAs  with moments along the $c$ axis.} Shown are the two components $\chi_{xx}^S$ and $\chi_{xy}^S$ of the SREE susceptibility tensor (top panels) and $\chi_{xx}^L$ and $\chi_{xy}^L$ of the OREE tensor (bottom panels), computed for scaled values of the SOC, as given in the panels. 
}
\label{fig:Fig3}
\end{figure}

\section{Discussion}

\subsection{Importance of spin-orbit interaction}

In his original work, Edelstein predicted an electrically induced out-of-equilibrium spin magnetization generated by Rashba SOC  \cite{Edelstein1990}. Here, without assuming any specific shape for the SOC, we find that, depending on the magnetic configuration, the symmetry of the induced magnetization can adopt  Rashba-like or Dresselhaus-like behaviors. Remarkably, we find that the previously neglected orbital polarization can in fact be much larger than the induced spin polarization.

The importance of SOC on the magnetoelectric susceptibilities can be accessed by reducing or switching off SOC in the calculations.
Doing so, we find that the spin Rashba-Edelstein effect computed without SOC completely vanishes; therefore, consistent with Edelstein's work \cite{Edelstein1990}, this is an intrinsic effect which occurs due to the broken inversion symmetry in the presence of SOC.
Surprisingly, however, for the orbital component our calculations without SOC give an unchanged, non-vanishing OREE response for the dominant off-diagonal tensor elements, as shown in  Fig.\ \ref{fig:Fig3}, bottom-right panel.
%
In Fig.\ \ref{fig:Fig3} (top panels) we show the computed SOC dependence of the $xx$ and $xy$ tensor elements of the SREE susceptibility of Mn1 in CuMnAs with antiferromagnetic moments along the $c$ axis.  These elements decrease linearly with a decreasing SOC.  
For the OREE,  in Fig.\ \ref{fig:Fig3} (bottom-right panel),
we find that the staggered components $\chi_{xy}^L$ {(and $\chi_{yx}^L$, not shown)} are  present even without SOC, and are not even changed by SOC strength which suggests that the leading off-diagonal term is independent of SOC. In contrast, without SOC the non-staggered OREE components $\chi_{xx}^L$ and also $\chi_{yy}^L$ and $\chi_{zz}^L$ (not shown) vanish, and these can consequently be identified as intrinsic SOC-related quantities. This observation is quite crucial and unexpected, since the staggered SREE components are generally believed to be at the origin of switching, in e.g.\ CuMnAs, and to be SOC related.  We find that the dominant nonrelativistic contribution is in the staggered OREE  components while smaller staggered spin components and non-staggered orbital components are generated by SOC. 

As yet we know little about the influence of the OREE for a magnetization switching event, but a cautioning remark is warranted. Although the OREE can be large, to act on the spin moments present in an AFM, it needs  to couple to these through spin-orbit interaction. Then, the overall torque on the antiferromagnetic spin moments will eventually be proportional to the SOC. 

To analyze the origin of the induced orbital polarizations, we observe that due to the staggered nature of the induced moments in Fig.\ \ref{fig:Fig3}, the sum of the induced orbital moments on all atoms in the unit cell cancels, but the contributions on individual atoms do not. There is thus an atomic orbital polarization present even without SOC.
The orbital angular momentum dynamics  induced by the applied potential $\hat{V}(t) = -e \bm{E}(t) \cdot \bm{\hat{r}}$ can be evaluated from the Heisenberg expression in a single-electron picture as
\begin{equation}
 \frac{d \bm{\hat{L}}^{ind}}{dt} =\frac{1}{i\hbar} \Big[ \bm{\hat{L}}^{ind}, \hat{V}(t) \Big] = \bm{\hat{r}} \times e \bm{E}(t) ,
\end{equation}
which is 
the quantum mechanical counter-part of the classical equation of motion for angular momentum, 
$\frac{d\bm{L}}{dt} = \bm{r} \times \bm{F}$, where $\bm{F}$ is 
 an externally applied force. In this picture, the electric field acts as a torque on the center of mass of the electrons on an atom. This does not require the interplay of SOC as the field couples directly to the position of the electrons and thereby affects the orbital momentum. Therefore, the OREE does not arises only from the small relativistic SOC, and sizeable effects might thus even be observed in systems with small SOC.
 
As a further remarkable aspect, we point out that the here-observed appearance of an orbital polarization in the unit cell in the absence of SOC is \textit{distinct} from other recent theoretical predictions of nonzero orbital textures \cite{Go2017,Hanke2017,Yoda2018}. Hanke \textit{et al.}\ showed that a nonzero static orbital moment can arise in the noncoplanar antiferromagnet $\gamma$-FeMn without SOC due to spin chirality \cite{Hanke2017}. Here, in the absence of spin chirality, we predict nonzero orbital moments that are present without SOC  when an applied electric field is present.  
Yoda \textit{et al.}\  proposed that in a chiral crystal  the solenoidal electron hopping motion could lead to an orbital magnetization in systems with time-reversal symmetry, in the sense of an orbital Edelstein effect \cite{Yoda2018}. Our here-computed induced orbital polarization is distinctly different, as it does not require  a chiral crystal symmetry.

\subsection{Frequency and magnetization dependence}

Our calculations predict sizable induced polarizations at finite frequencies, which raises the question whether electric field driven magnetic moment switching could be achieved at high frequencies. It is well known that 
time-dependent magnetic fields cannot drive fast spin dynamics of ferromagnets in the optical regime because the magnetic permeability $\mu(\omega)$ decays quickly as $\omega$ increases to the infrared region \cite{Kittel1946}.
The situation is however entirely different for the SREE. The magnetic permeability is due to a magnetic field $\bm H$ that acts on the spin through the Zeeman interaction in the Hamiltonian, $\mu_B \hat{\bm{\sigma}} \cdot \bm{H}(t)$, whereas for the SREE the electric field couples to the charge, $-e \hat{\bm{r}} \cdot \bm{E}(t) $.  The electric charges can indeed follow the rapidly changing $E$-field, implying that an equally fast magnetic response can be anticipated. Due to their electrical origin, the induced magnetizations can be driven at petahertz frequencies, thus opening for potential  routes to achieve petahertz spintronics.

In the DC limit, $\omega =0$, the real part of the REE is nonzero and its imaginary part vanishes exactly. At finite frequencies, both the real and imaginary parts of the tensor components can be nonzero. The nonzero imaginary REE susceptibility has a specific influence on the evolving magnetization dynamics. For a given driving electric field $\bm{E}(t)$, the induced spin polarization $\delta\bm{M}^S (t)$ can  be retrieved from a Fourier transform of $ \delta\bm{M}^S (\omega) = \mbox{\large$\chi$}^S(\omega) \bm{E}(\omega)$. The induced spin polarization follows the driving field, but it has a phase difference due to the imaginary SREE susceptibility. An equivalent relation holds for the orbital polarization.
The induced spin and orbital polarizations at a frequency $\omega$ will thus still provide staggered torques on the existing static moments, but these torques will alternate with time. A major question is then how fast the switching of the static moments can proceed, whether this can be pushed to the PHz regime. Recent experiments demonstrated that switching of CuMnAs is possible at THz frequencies \cite{Olejnik2018}. Potentially, on account of the above, the switching could thus be even faster in antiferromagnets, in particular when the torques could be enhanced, but the boundaries on the switching speed are as yet unexplored.

To verify whether the SREE and/or OREE can be at the origin of ultrafast switching, and what the intrinsic frequency limit is, one should perform atomistic spin-dynamics simulations. The inclusion of both induced spin and orbital magnetic moments would notably be required to achieve the full picture. This in turn would necessitate the handling of two coupled moments per atom, as was done recently for the dynamics of the $4f$ and $5d$ moments on gadolinium \cite{Wienholdt2013}. Such spin-dynamics simulations could clarify as well the role of the non-staggered, homogeneous components for the switching and the influence of Joule heating, inherently present in all experiments. It was shown recently that Joule heating plays an essential role as it drastically decreases the required switching field and enhances the spin-orbit torque efficiency \cite{Li2018}. Also for Mn$_2$Au it was lately concluded that Joule heating can provide a sufficient thermal activation for switching processes \cite{Meinert2018}.
Lastly, it should be emphasized that the switching dynamics of an antiferromagnet is distinct from that of a ferromagnet, since the magnetization dynamics of an antiferromagnet is described by a second-order differential equation, which contains a magnetic inertia term for the spins \cite{Kim2017a,Kimel2009,Mondal2017}. This  antiferromagnetic inertia can provide an important stimulus  for the switching, because, even after the pulse is switched off,  the already induced torques will act for a longer time as drivers of the dynamics. 

\section{Conclusions}

Switching in antiferromagnets is believed to be due to locally staggered spin-orbit fields that drive opposite dynamics of moments on the two AFM sublattices  \cite{Zelezny2014,Grzybowski2017,Bodnar2018,Olejnik2018,Wadley2016,Godinho2018,Meinert2018,Olejnik2017}. 
Our investigation strongly supports that the Rashba-Edelstein effect is an excellent candidate to explain the microscopic origin of such staggered fields. Beyond this, we report several surprising discoveries: first, there exists a significant orbital Rashba-Edelstein effect that can be much larger than the spin Rashba-Edelstein effect.
Second, we find that 
 there exists not only staggered but also non-staggered components to the REE tensors. In both CuMnAs and Mn$_2$Au, we find that the staggered response is strongest. {This causes a locking of the orbital momentum perpendicular to the applied field.} 
 
 Computing the symmetry of the induced polarizations with respect to an in-plane electric field, we find that these can have a Rashba-like or a Dresselhaus-like character and that these characters can {in general} be distinct for the induced spin and orbital polarizations; for example, a {Dresselhaus-like} symmetry  for the SREE and a {Rashba} symmetry  for the OREE of CuMnAs with in-plane AFM moments. As a consequence, the spin and orbital fields can enhance each other or cancel each other, i.e.,  act in a cooperative or a non-cooperative way for switching of the sublattice magnetizations. 

The most surprising part of this work is undoubtedly the strong induced orbital polarization, which can be much larger than the induced spin dipole magnetization. The nonequilibrium orbital polarization is notably even present  in the absence of spin-orbit interaction. This implies that it does not arise from a small relativistic effect, but has a more fundamental,  nonrelativistic origin. While our focus here has been on the two antiferromagnets that are of current interest for antiferromagnetic spintronics, the large dominant orbital fields could gain importance in the {emerging} field of spinorbitronics \cite{Manchon2017}. As the induced spin and orbital polarizations originate from the coupling of the electric field to the electron charges, these induced polarizations can moreover be driven at high frequencies, opening prospects for achieving spintronics at petahertz frequencies.

Lastly, on a more general note, the here-developed general \textit{ab initio} framework can be employed for the study of nonequilibrium electric-field induced polarizations in a wide range of materials, as e.g.\ bulk compounds and metal/ferromagnet or metal/antiferromagnet interfaces. While bulk materials can already display rich spin-orbit-related physics,  interfaces of a heavy metal with a magnetic layer, where the SOC is increased at the interface, can feature an enhanced Rashba-Edelstein effect that can e.g.\ be utilized to control the spin orientation in the magnetic layer \cite{Manchon2015}. 
 {Our} \textit{ab initio} theory {framework} can provide a materials' specific understanding of the  mechanisms behind {electrical} spin control and lead to the design of suitable interfaces for future spintronics applications.

\section{Acknowledgment}
This work has been supported by 
the Swedish Research Council (VR), the K.\ and A.\ Wallenberg Foundation (grant No.\ 2015.0060), the European Union's Horizon2020 Research and Innovation Programme under grant agreement No.\ 737709 (FEMTOTERABYTE), 
and the Swedish National  Infrastructure for Computing (SNIC).	 
The calculations were performed at the PDC Center for High Performance Computing and the Uppsala Multidisciplinary Center for Advanced Computational Science (UPPMAX). 

\section{Appendix}

To evaluate the frequency-dependent SREE and OREE tensors, we adopt the DFT formalism as implemented in the full-potential linearized augmented plane-wave (FLAPW) all-electron code WIEN2k \cite{Blaha2018}. The momentum matrix elements are computed through the WIEN2k package \cite{Ambrosch-Draxl2006} while we use our own implementation for the spin and orbital momentum matrix elements. In all of our calculations, we use the PBE-GGA exchange correlation potential \cite{Perdew1996}. The broadening $\hbar\tau^{-1}$ is set to $0.41$ eV for both intra- and inter-band transitions which was found to give realistic results for the (spin-orbit related) magneto-optical properties of metallic systems \cite{Oppeneer2001}. The spin and orbital responses to the electric field {are} computed over a whole range of frequency, i.e., we do not restrict our formalism to static electric fields ($\omega=0$). Both the real and imaginary parts of the SREE and OREE susceptibility tensors are computed. Our relativistic DFT calculations include spin-orbit interaction consistent with Dirac theory, without resorting to any specific form of SOC such as Rashba or Dresselhaus.


For CuMnAs, the product between the smallest muffin-tin radius ${R}_{\text{MT}}$ and the largest reciprocal vector ${K}_{\text{max}}$ was ${R}_{\text{MT}} \times {K}_{\text{max}}=7.5$. The self-consistent spin-polarized density was computed using a $25\times25\times15$ Monkhorst-Pack grid \cite{Monkhorst1976}. The REE tensors were then computed using a $43\times43\times26$ $k$-point grid.
For Mn$_\text{2}$Au, the product between the smallest muffin-tin radius ${R}_{\text{MT}}$ and the largest reciprocal vector ${K}_{\text{max}}$ was ${R}_{\text{MT}}\times {K}_{\text{max}}=8$. The self-consistent spin-polarized density was computed using a $29\times29\times11$ Monkhorst-Pack grid \cite{Monkhorst1976}. The REE tensors were computed on a $42\times42\times16$ $k$-point grid. {The numerical convergence of the REE tensor elements was checked by using much denser $k$-point grids.}
We adopted the experimental lattice constants in our calculations \cite{Barthem2013,Wadley2013}.

\bibliographystyle{apsrev4-1}

\begin{thebibliography}{56}%
\makeatletter
\providecommand \@ifxundefined [1]{%
 \@ifx{#1\undefined}
}%
\providecommand \@ifnum [1]{%
 \ifnum #1\expandafter \@firstoftwo
 \else \expandafter \@secondoftwo
 \fi
}%
\providecommand \@ifx [1]{%
 \ifx #1\expandafter \@firstoftwo
 \else \expandafter \@secondoftwo
 \fi
}%
\providecommand \natexlab [1]{#1}%
\providecommand \enquote  [1]{``#1''}%
\providecommand \bibnamefont  [1]{#1}%
\providecommand \bibfnamefont [1]{#1}%
\providecommand \citenamefont [1]{#1}%
\providecommand \href@noop [0]{\@secondoftwo}%
\providecommand \href [0]{\begingroup \@sanitize@url \@href}%
\providecommand \@href[1]{\@@startlink{#1}\@@href}%
\providecommand \@@href[1]{\endgroup#1\@@endlink}%
\providecommand \@sanitize@url [0]{\catcode `\\12\catcode `\$12\catcode
  `\&12\catcode `\#12\catcode `\^12\catcode `\_12\catcode `\%12\relax}%
\providecommand \@@startlink[1]{}%
\providecommand \@@endlink[0]{}%
\providecommand \url  [0]{\begingroup\@sanitize@url \@url }%
\providecommand \@url [1]{\endgroup\@href {#1}{\urlprefix }}%
\providecommand \urlprefix  [0]{URL }%
\providecommand \Eprint [0]{\href }%
\providecommand \doibase [0]{http://dx.doi.org/}%
\providecommand \selectlanguage [0]{\@gobble}%
\providecommand \bibinfo  [0]{\@secondoftwo}%
\providecommand \bibfield  [0]{\@secondoftwo}%
\providecommand \translation [1]{[#1]}%
\providecommand \BibitemOpen [0]{}%
\providecommand \bibitemStop [0]{}%
\providecommand \bibitemNoStop [0]{.\EOS\space}%
\providecommand \EOS [0]{\spacefactor3000\relax}%
\providecommand \BibitemShut  [1]{\csname bibitem#1\endcsname}%
\let\auto@bib@innerbib\@empty
\bibitem [{\citenamefont {{\v{Z}}uti{\'{c}}}\ \emph {et~al.}(2004)\citenamefont
  {{\v{Z}}uti{\'{c}}}, \citenamefont {Fabian},\ and\ \citenamefont {{Das
  Sarma}}}]{Zutic2004}%
  \BibitemOpen
  \bibfield  {author} {\bibinfo {author} {\bibfnamefont {I.}~\bibnamefont
  {{\v{Z}}uti{\'{c}}}}, \bibinfo {author} {\bibfnamefont {J.}~\bibnamefont
  {Fabian}}, \ and\ \bibinfo {author} {\bibfnamefont {S.}~\bibnamefont {{Das
  Sarma}}},\ }\href {\doibase 10.1103/RevModPhys.76.323} {\bibfield  {journal}
  {\bibinfo  {journal} {Rev. Mod. Phys.}\ }\textbf {\bibinfo {volume} {76}},\
  \bibinfo {pages} {323} (\bibinfo {year} {2004})}\BibitemShut {NoStop}%
\bibitem [{\citenamefont {Brataas}\ \emph {et~al.}(2012)\citenamefont
  {Brataas}, \citenamefont {Kent},\ and\ \citenamefont {Ohno}}]{Brataas2012}%
  \BibitemOpen
  \bibfield  {author} {\bibinfo {author} {\bibfnamefont {A.}~\bibnamefont
  {Brataas}}, \bibinfo {author} {\bibfnamefont {A.~D.}\ \bibnamefont {Kent}}, \
  and\ \bibinfo {author} {\bibfnamefont {H.}~\bibnamefont {Ohno}},\ }\href@noop
  {} {\bibfield  {journal} {\bibinfo  {journal} {Nat. Mater.}\ }\textbf
  {\bibinfo {volume} {11}},\ \bibinfo {pages} {372} (\bibinfo {year}
  {2012})}\BibitemShut {NoStop}%
\bibitem [{\citenamefont {Hellman}\ \emph {et~al.}(2017)\citenamefont
  {Hellman}, \citenamefont {Hoffmann}, \citenamefont {Tserkovnyak},
  \citenamefont {Beach}, \citenamefont {Fullerton}, \citenamefont {Leighton},
  \citenamefont {MacDonald}, \citenamefont {Ralph}, \citenamefont {Arena},
  \citenamefont {D\"urr}, \citenamefont {Fischer}, \citenamefont {Grollier},
  \citenamefont {Heremans}, \citenamefont {Jungwirth}, \citenamefont {Kimel},
  \citenamefont {Koopmans}, \citenamefont {Krivorotov}, \citenamefont {May},
  \citenamefont {Petford-Long}, \citenamefont {Rondinelli}, \citenamefont
  {Samarth}, \citenamefont {Schuller}, \citenamefont {Slavin}, \citenamefont
  {Stiles}, \citenamefont {Tchernyshyov}, \citenamefont {Thiaville},\ and\
  \citenamefont {Zink}}]{Hellman2017}%
  \BibitemOpen
  \bibfield  {author} {\bibinfo {author} {\bibfnamefont {F.}~\bibnamefont
  {Hellman}}, \bibinfo {author} {\bibfnamefont {A.}~\bibnamefont {Hoffmann}},
  \bibinfo {author} {\bibfnamefont {Y.}~\bibnamefont {Tserkovnyak}}, \bibinfo
  {author} {\bibfnamefont {G.~S.~D.}\ \bibnamefont {Beach}}, \bibinfo {author}
  {\bibfnamefont {E.~E.}\ \bibnamefont {Fullerton}}, \bibinfo {author}
  {\bibfnamefont {C.}~\bibnamefont {Leighton}}, \bibinfo {author}
  {\bibfnamefont {A.~H.}\ \bibnamefont {MacDonald}}, \bibinfo {author}
  {\bibfnamefont {D.~C.}\ \bibnamefont {Ralph}}, \bibinfo {author}
  {\bibfnamefont {D.~A.}\ \bibnamefont {Arena}}, \bibinfo {author}
  {\bibfnamefont {H.~A.}\ \bibnamefont {D\"urr}}, \bibinfo {author}
  {\bibfnamefont {P.}~\bibnamefont {Fischer}}, \bibinfo {author} {\bibfnamefont
  {J.}~\bibnamefont {Grollier}}, \bibinfo {author} {\bibfnamefont {J.~P.}\
  \bibnamefont {Heremans}}, \bibinfo {author} {\bibfnamefont {T.}~\bibnamefont
  {Jungwirth}}, \bibinfo {author} {\bibfnamefont {A.~V.}\ \bibnamefont
  {Kimel}}, \bibinfo {author} {\bibfnamefont {B.}~\bibnamefont {Koopmans}},
  \bibinfo {author} {\bibfnamefont {I.~N.}\ \bibnamefont {Krivorotov}},
  \bibinfo {author} {\bibfnamefont {S.~J.}\ \bibnamefont {May}}, \bibinfo
  {author} {\bibfnamefont {A.~K.}\ \bibnamefont {Petford-Long}}, \bibinfo
  {author} {\bibfnamefont {J.~M.}\ \bibnamefont {Rondinelli}}, \bibinfo
  {author} {\bibfnamefont {N.}~\bibnamefont {Samarth}}, \bibinfo {author}
  {\bibfnamefont {I.~K.}\ \bibnamefont {Schuller}}, \bibinfo {author}
  {\bibfnamefont {A.~N.}\ \bibnamefont {Slavin}}, \bibinfo {author}
  {\bibfnamefont {M.~D.}\ \bibnamefont {Stiles}}, \bibinfo {author}
  {\bibfnamefont {O.}~\bibnamefont {Tchernyshyov}}, \bibinfo {author}
  {\bibfnamefont {A.}~\bibnamefont {Thiaville}}, \ and\ \bibinfo {author}
  {\bibfnamefont {B.~L.}\ \bibnamefont {Zink}},\ }\href {\doibase
  10.1103/RevModPhys.89.025006} {\bibfield  {journal} {\bibinfo  {journal}
  {Rev. Mod. Phys.}\ }\textbf {\bibinfo {volume} {89}},\ \bibinfo {pages}
  {025006} (\bibinfo {year} {2017})}\BibitemShut {NoStop}%
\bibitem [{\citenamefont {Jungwirth}\ \emph {et~al.}(2016)\citenamefont
  {Jungwirth}, \citenamefont {Marti}, \citenamefont {Wadley},\ and\
  \citenamefont {Wunderlich}}]{Jungwirth2016}%
  \BibitemOpen
  \bibfield  {author} {\bibinfo {author} {\bibfnamefont {T.}~\bibnamefont
  {Jungwirth}}, \bibinfo {author} {\bibfnamefont {X.}~\bibnamefont {Marti}},
  \bibinfo {author} {\bibfnamefont {P.}~\bibnamefont {Wadley}}, \ and\ \bibinfo
  {author} {\bibfnamefont {J.}~\bibnamefont {Wunderlich}},\ }\href@noop {}
  {\bibfield  {journal} {\bibinfo  {journal} {Nat. Nanotechn.}\ }\textbf
  {\bibinfo {volume} {11}},\ \bibinfo {pages} {231} (\bibinfo {year}
  {2016})}\BibitemShut {NoStop}%
\bibitem [{\citenamefont {N{\v{e}}mec}\ \emph {et~al.}(2018)\citenamefont
  {N{\v{e}}mec}, \citenamefont {Fiebig}, \citenamefont {Kampfrath},\ and\
  \citenamefont {Kimel}}]{Nemec2018}%
  \BibitemOpen
  \bibfield  {author} {\bibinfo {author} {\bibfnamefont {P.}~\bibnamefont
  {N{\v{e}}mec}}, \bibinfo {author} {\bibfnamefont {M.}~\bibnamefont {Fiebig}},
  \bibinfo {author} {\bibfnamefont {T.}~\bibnamefont {Kampfrath}}, \ and\
  \bibinfo {author} {\bibfnamefont {A.~V.}\ \bibnamefont {Kimel}},\ }\href@noop
  {} {\bibfield  {journal} {\bibinfo  {journal} {Nat. Phys.}\ }\textbf
  {\bibinfo {volume} {14}},\ \bibinfo {pages} {229} (\bibinfo {year}
  {2018})}\BibitemShut {NoStop}%
\bibitem [{\citenamefont {Baltz}\ \emph {et~al.}(2018)\citenamefont {Baltz},
  \citenamefont {Manchon}, \citenamefont {Tsoi}, \citenamefont {Moriyama},
  \citenamefont {Ono},\ and\ \citenamefont {Tserkovnyak}}]{Baltz2018}%
  \BibitemOpen
  \bibfield  {author} {\bibinfo {author} {\bibfnamefont {V.}~\bibnamefont
  {Baltz}}, \bibinfo {author} {\bibfnamefont {A.}~\bibnamefont {Manchon}},
  \bibinfo {author} {\bibfnamefont {M.}~\bibnamefont {Tsoi}}, \bibinfo {author}
  {\bibfnamefont {T.}~\bibnamefont {Moriyama}}, \bibinfo {author}
  {\bibfnamefont {T.}~\bibnamefont {Ono}}, \ and\ \bibinfo {author}
  {\bibfnamefont {Y.}~\bibnamefont {Tserkovnyak}},\ }\href {\doibase
  10.1103/RevModPhys.90.015005} {\bibfield  {journal} {\bibinfo  {journal}
  {Rev. Mod. Phys.}\ }\textbf {\bibinfo {volume} {90}},\ \bibinfo {pages}
  {015005} (\bibinfo {year} {2018})}\BibitemShut {NoStop}%
\bibitem [{\citenamefont {M{\'{a}}ca}\ \emph {et~al.}(2012)\citenamefont
  {M{\'{a}}ca}, \citenamefont {Ma{\v{s}}ek}, \citenamefont {Stelmakhovych},
  \citenamefont {Mart{\'{i}}}, \citenamefont {Reichlov{\'{a}}}, \citenamefont
  {Uhl{\'{i}}?ov{\'{a}}}, \citenamefont {Beran}, \citenamefont {Wadley},
  \citenamefont {Nov{\'{a}}k},\ and\ \citenamefont {Jungwirth}}]{Maca2012}%
  \BibitemOpen
  \bibfield  {author} {\bibinfo {author} {\bibfnamefont {F.}~\bibnamefont
  {M{\'{a}}ca}}, \bibinfo {author} {\bibfnamefont {J.}~\bibnamefont
  {Ma{\v{s}}ek}}, \bibinfo {author} {\bibfnamefont {O.}~\bibnamefont
  {Stelmakhovych}}, \bibinfo {author} {\bibfnamefont {X.}~\bibnamefont
  {Mart{\'{i}}}}, \bibinfo {author} {\bibfnamefont {H.}~\bibnamefont
  {Reichlov{\'{a}}}}, \bibinfo {author} {\bibfnamefont {K.}~\bibnamefont
  {Uhl{\'{i}}?ov{\'{a}}}}, \bibinfo {author} {\bibfnamefont {P.}~\bibnamefont
  {Beran}}, \bibinfo {author} {\bibfnamefont {P.}~\bibnamefont {Wadley}},
  \bibinfo {author} {\bibfnamefont {V.}~\bibnamefont {Nov{\'{a}}k}}, \ and\
  \bibinfo {author} {\bibfnamefont {T.}~\bibnamefont {Jungwirth}},\ }\href
  {\doibase 10.1016/J.JMMM.2011.12.017} {\bibfield  {journal} {\bibinfo
  {journal} {J. Magn. Magn. Mater.}\ }\textbf {\bibinfo {volume} {324}},\
  \bibinfo {pages} {1606} (\bibinfo {year} {2012})}\BibitemShut {NoStop}%
\bibitem [{\citenamefont {Yamaoka}(1974)}]{Yamaoka1974}%
  \BibitemOpen
  \bibfield  {author} {\bibinfo {author} {\bibfnamefont {T.}~\bibnamefont
  {Yamaoka}},\ }\href {\doibase 10.1143/JPSJ.36.445} {\bibfield  {journal}
  {\bibinfo  {journal} {J. Phys. Soc. Jpn.}\ }\textbf {\bibinfo {volume}
  {36}},\ \bibinfo {pages} {445} (\bibinfo {year} {1974})}\BibitemShut
  {NoStop}%
\bibitem [{\citenamefont {Barthem}\ \emph {et~al.}(2013)\citenamefont
  {Barthem}, \citenamefont {Colin}, \citenamefont {Mayaffre}, \citenamefont
  {Julien},\ and\ \citenamefont {Givord}}]{Barthem2013}%
  \BibitemOpen
  \bibfield  {author} {\bibinfo {author} {\bibfnamefont {V.~M. T.~S.}\
  \bibnamefont {Barthem}}, \bibinfo {author} {\bibfnamefont {C.~V.}\
  \bibnamefont {Colin}}, \bibinfo {author} {\bibfnamefont {H.}~\bibnamefont
  {Mayaffre}}, \bibinfo {author} {\bibfnamefont {M.~H.}\ \bibnamefont
  {Julien}}, \ and\ \bibinfo {author} {\bibfnamefont {D.}~\bibnamefont
  {Givord}},\ }\href {\doibase 10.1038/ncomms3892} {\bibfield  {journal}
  {\bibinfo  {journal} {Nat. Commun.}\ }\textbf {\bibinfo {volume} {4}},\
  \bibinfo {pages} {2892} (\bibinfo {year} {2013})}\BibitemShut {NoStop}%
\bibitem [{\citenamefont {Kimel}\ \emph {et~al.}(2004)\citenamefont {Kimel},
  \citenamefont {Kirilyuk}, \citenamefont {Tsvetkov}, \citenamefont {Pisarev},\
  and\ \citenamefont {Rasing}}]{Kimel2004}%
  \BibitemOpen
  \bibfield  {author} {\bibinfo {author} {\bibfnamefont {A.~V.}\ \bibnamefont
  {Kimel}}, \bibinfo {author} {\bibfnamefont {A.}~\bibnamefont {Kirilyuk}},
  \bibinfo {author} {\bibfnamefont {A.}~\bibnamefont {Tsvetkov}}, \bibinfo
  {author} {\bibfnamefont {R.~V.}\ \bibnamefont {Pisarev}}, \ and\ \bibinfo
  {author} {\bibfnamefont {T.}~\bibnamefont {Rasing}},\ }\href {\doibase
  10.1038/nature02659} {\bibfield  {journal} {\bibinfo  {journal} {Nature}\
  }\textbf {\bibinfo {volume} {429}},\ \bibinfo {pages} {850} (\bibinfo {year}
  {2004})}\BibitemShut {NoStop}%
\bibitem [{\citenamefont {Fiebig}\ \emph {et~al.}(2008)\citenamefont {Fiebig},
  \citenamefont {Duong}, \citenamefont {Satoh}, \citenamefont {{Van Aken}},
  \citenamefont {Miyano}, \citenamefont {Tomioka},\ and\ \citenamefont
  {Tokura}}]{Fiebig2008}%
  \BibitemOpen
  \bibfield  {author} {\bibinfo {author} {\bibfnamefont {M.}~\bibnamefont
  {Fiebig}}, \bibinfo {author} {\bibfnamefont {N.~P.}\ \bibnamefont {Duong}},
  \bibinfo {author} {\bibfnamefont {T.}~\bibnamefont {Satoh}}, \bibinfo
  {author} {\bibfnamefont {B.~B.}\ \bibnamefont {{Van Aken}}}, \bibinfo
  {author} {\bibfnamefont {K.}~\bibnamefont {Miyano}}, \bibinfo {author}
  {\bibfnamefont {Y.}~\bibnamefont {Tomioka}}, \ and\ \bibinfo {author}
  {\bibfnamefont {Y.}~\bibnamefont {Tokura}},\ }\href {\doibase
  10.1088/0022-3727/41/16/164005} {\bibfield  {journal} {\bibinfo  {journal}
  {J. Phys. D: Appl. Phys.}\ }\textbf {\bibinfo {volume} {41}},\ \bibinfo
  {pages} {164005} (\bibinfo {year} {2008})}\BibitemShut {NoStop}%
\bibitem [{\citenamefont {Kampfrath}\ \emph {et~al.}(2010)\citenamefont
  {Kampfrath}, \citenamefont {Sell}, \citenamefont {Klatt}, \citenamefont
  {Pashkin}, \citenamefont {M{\"{a}}hrlein}, \citenamefont {Dekorsy},
  \citenamefont {Wolf}, \citenamefont {Fiebig}, \citenamefont {Leitenstorfer},\
  and\ \citenamefont {Huber}}]{Kampfrath2010}%
  \BibitemOpen
  \bibfield  {author} {\bibinfo {author} {\bibfnamefont {T.}~\bibnamefont
  {Kampfrath}}, \bibinfo {author} {\bibfnamefont {A.}~\bibnamefont {Sell}},
  \bibinfo {author} {\bibfnamefont {G.}~\bibnamefont {Klatt}}, \bibinfo
  {author} {\bibfnamefont {A.}~\bibnamefont {Pashkin}}, \bibinfo {author}
  {\bibfnamefont {S.}~\bibnamefont {M{\"{a}}hrlein}}, \bibinfo {author}
  {\bibfnamefont {T.}~\bibnamefont {Dekorsy}}, \bibinfo {author} {\bibfnamefont
  {M.}~\bibnamefont {Wolf}}, \bibinfo {author} {\bibfnamefont {M.}~\bibnamefont
  {Fiebig}}, \bibinfo {author} {\bibfnamefont {A.}~\bibnamefont
  {Leitenstorfer}}, \ and\ \bibinfo {author} {\bibfnamefont {R.}~\bibnamefont
  {Huber}},\ }\href {\doibase 10.1038/nphoton.2010.259} {\bibfield  {journal}
  {\bibinfo  {journal} {Nat. Photon.}\ }\textbf {\bibinfo {volume} {5}},\
  \bibinfo {pages} {31} (\bibinfo {year} {2010})}\BibitemShut {NoStop}%
\bibitem [{\citenamefont {Pashkin}\ \emph {et~al.}(2013)\citenamefont
  {Pashkin}, \citenamefont {Sell}, \citenamefont {Kampfrath},\ and\
  \citenamefont {Huber}}]{Pashkin2013}%
  \BibitemOpen
  \bibfield  {author} {\bibinfo {author} {\bibfnamefont {A.}~\bibnamefont
  {Pashkin}}, \bibinfo {author} {\bibfnamefont {A.}~\bibnamefont {Sell}},
  \bibinfo {author} {\bibfnamefont {T.}~\bibnamefont {Kampfrath}}, \ and\
  \bibinfo {author} {\bibfnamefont {R.}~\bibnamefont {Huber}},\ }\href
  {\doibase 10.1088/1367-2630/15/6/065003} {\bibfield  {journal} {\bibinfo
  {journal} {New J. Phys.}\ }\textbf {\bibinfo {volume} {15}},\ \bibinfo
  {pages} {065003} (\bibinfo {year} {2013})}\BibitemShut {NoStop}%
\bibitem [{\citenamefont {Kirilyuk}\ \emph {et~al.}(2010)\citenamefont
  {Kirilyuk}, \citenamefont {Kimel},\ and\ \citenamefont
  {Rasing}}]{Kirilyuk2010}%
  \BibitemOpen
  \bibfield  {author} {\bibinfo {author} {\bibfnamefont {A.}~\bibnamefont
  {Kirilyuk}}, \bibinfo {author} {\bibfnamefont {A.~V.}\ \bibnamefont {Kimel}},
  \ and\ \bibinfo {author} {\bibfnamefont {T.}~\bibnamefont {Rasing}},\ }\href
  {https://journals.aps.org/rmp/pdf/10.1103/RevModPhys.82.2731} {\bibfield
  {journal} {\bibinfo  {journal} {Rev. Mod. Phys.}\ }\textbf {\bibinfo {volume}
  {82}},\ \bibinfo {pages} {2731} (\bibinfo {year} {2010})}\BibitemShut
  {NoStop}%
\bibitem [{\citenamefont {Kato}\ \emph
  {et~al.}(2004{\natexlab{a}})\citenamefont {Kato}, \citenamefont {Myers},
  \citenamefont {Gossard},\ and\ \citenamefont {Awschalom}}]{Kato2004a}%
  \BibitemOpen
  \bibfield  {author} {\bibinfo {author} {\bibfnamefont {Y.~K.}\ \bibnamefont
  {Kato}}, \bibinfo {author} {\bibfnamefont {R.~C.}\ \bibnamefont {Myers}},
  \bibinfo {author} {\bibfnamefont {A.~C.}\ \bibnamefont {Gossard}}, \ and\
  \bibinfo {author} {\bibfnamefont {D.~D.}\ \bibnamefont {Awschalom}},\ }\href
  {\doibase 10.1126/science.1105514} {\bibfield  {journal} {\bibinfo  {journal}
  {Science}\ }\textbf {\bibinfo {volume} {306}},\ \bibinfo {pages} {1910}
  (\bibinfo {year} {2004}{\natexlab{a}})}\BibitemShut {NoStop}%
\bibitem [{\citenamefont {Sinova}\ \emph {et~al.}(2015)\citenamefont {Sinova},
  \citenamefont {Valenzuela}, \citenamefont {Wunderlich}, \citenamefont
  {Back},\ and\ \citenamefont {Jungwirth}}]{Sinova2015}%
  \BibitemOpen
  \bibfield  {author} {\bibinfo {author} {\bibfnamefont {J.}~\bibnamefont
  {Sinova}}, \bibinfo {author} {\bibfnamefont {S.~O.}\ \bibnamefont
  {Valenzuela}}, \bibinfo {author} {\bibfnamefont {J.}~\bibnamefont
  {Wunderlich}}, \bibinfo {author} {\bibfnamefont {C.~H.}\ \bibnamefont
  {Back}}, \ and\ \bibinfo {author} {\bibfnamefont {T.}~\bibnamefont
  {Jungwirth}},\ }\href {\doibase 10.1103/RevModPhys.87.1213} {\bibfield
  {journal} {\bibinfo  {journal} {Rev. Mod. Phys.}\ }\textbf {\bibinfo {volume}
  {87}},\ \bibinfo {pages} {1213} (\bibinfo {year} {2015})}\BibitemShut
  {NoStop}%
\bibitem [{\citenamefont {Miron}\ \emph {et~al.}(2011)\citenamefont {Miron},
  \citenamefont {Garello}, \citenamefont {Gaudin}, \citenamefont {Zermatten},
  \citenamefont {Costache}, \citenamefont {Auffret}, \citenamefont {Bandiera},
  \citenamefont {Rodmacq}, \citenamefont {Schuhl},\ and\ \citenamefont
  {Gambardella}}]{Miron2011}%
  \BibitemOpen
  \bibfield  {author} {\bibinfo {author} {\bibfnamefont {I.~M.}\ \bibnamefont
  {Miron}}, \bibinfo {author} {\bibfnamefont {K.}~\bibnamefont {Garello}},
  \bibinfo {author} {\bibfnamefont {G.}~\bibnamefont {Gaudin}}, \bibinfo
  {author} {\bibfnamefont {P.-J.}\ \bibnamefont {Zermatten}}, \bibinfo {author}
  {\bibfnamefont {M.~V.}\ \bibnamefont {Costache}}, \bibinfo {author}
  {\bibfnamefont {S.}~\bibnamefont {Auffret}}, \bibinfo {author} {\bibfnamefont
  {S.}~\bibnamefont {Bandiera}}, \bibinfo {author} {\bibfnamefont
  {B.}~\bibnamefont {Rodmacq}}, \bibinfo {author} {\bibfnamefont
  {A.}~\bibnamefont {Schuhl}}, \ and\ \bibinfo {author} {\bibfnamefont
  {P.}~\bibnamefont {Gambardella}},\ }\href@noop {} {\bibfield  {journal}
  {\bibinfo  {journal} {Nature}\ }\textbf {\bibinfo {volume} {476}},\ \bibinfo
  {pages} {189} (\bibinfo {year} {2011})}\BibitemShut {NoStop}%
\bibitem [{\citenamefont {Liu}\ \emph {et~al.}(2012)\citenamefont {Liu},
  \citenamefont {Pai}, \citenamefont {Li}, \citenamefont {Tseng}, \citenamefont
  {Ralph},\ and\ \citenamefont {Buhrman}}]{Liu2012}%
  \BibitemOpen
  \bibfield  {author} {\bibinfo {author} {\bibfnamefont {L.}~\bibnamefont
  {Liu}}, \bibinfo {author} {\bibfnamefont {C.-F.}\ \bibnamefont {Pai}},
  \bibinfo {author} {\bibfnamefont {Y.}~\bibnamefont {Li}}, \bibinfo {author}
  {\bibfnamefont {H.~W.}\ \bibnamefont {Tseng}}, \bibinfo {author}
  {\bibfnamefont {D.~C.}\ \bibnamefont {Ralph}}, \ and\ \bibinfo {author}
  {\bibfnamefont {R.~A.}\ \bibnamefont {Buhrman}},\ }\href@noop {} {\bibfield
  {journal} {\bibinfo  {journal} {Science}\ }\textbf {\bibinfo {volume}
  {336}},\ \bibinfo {pages} {555} (\bibinfo {year} {2012})}\BibitemShut
  {NoStop}%
\bibitem [{\citenamefont {Garello}\ \emph {et~al.}(2013)\citenamefont
  {Garello}, \citenamefont {Miron}, \citenamefont {Avci}, \citenamefont
  {Freimuth}, \citenamefont {Mokrousov}, \citenamefont {Bl{\"u}gel},
  \citenamefont {Auffret}, \citenamefont {Boulle}, \citenamefont {Gaudin},\
  and\ \citenamefont {Gambardella}}]{Garello2013}%
  \BibitemOpen
  \bibfield  {author} {\bibinfo {author} {\bibfnamefont {K.}~\bibnamefont
  {Garello}}, \bibinfo {author} {\bibfnamefont {I.~M.}\ \bibnamefont {Miron}},
  \bibinfo {author} {\bibfnamefont {C.~O.}\ \bibnamefont {Avci}}, \bibinfo
  {author} {\bibfnamefont {F.}~\bibnamefont {Freimuth}}, \bibinfo {author}
  {\bibfnamefont {Y.}~\bibnamefont {Mokrousov}}, \bibinfo {author}
  {\bibfnamefont {S.}~\bibnamefont {Bl{\"u}gel}}, \bibinfo {author}
  {\bibfnamefont {S.}~\bibnamefont {Auffret}}, \bibinfo {author} {\bibfnamefont
  {O.}~\bibnamefont {Boulle}}, \bibinfo {author} {\bibfnamefont
  {G.}~\bibnamefont {Gaudin}}, \ and\ \bibinfo {author} {\bibfnamefont
  {P.}~\bibnamefont {Gambardella}},\ }\href@noop {} {\bibfield  {journal}
  {\bibinfo  {journal} {Nat. Nanotechnol.}\ }\textbf {\bibinfo {volume} {8}},\
  \bibinfo {pages} {587} (\bibinfo {year} {2013})}\BibitemShut {NoStop}%
\bibitem [{\citenamefont {Baumgartner}\ \emph {et~al.}(2017)\citenamefont
  {Baumgartner}, \citenamefont {Garello}, \citenamefont {Mendil}, \citenamefont
  {Avci}, \citenamefont {Grimaldi}, \citenamefont {Murer}, \citenamefont
  {Feng}, \citenamefont {Gabureac}, \citenamefont {Stamm}, \citenamefont
  {Acremann}, \citenamefont {Finizio}, \citenamefont {Wintz}, \citenamefont
  {Raabe},\ and\ \citenamefont {Gambardella}}]{Baumgartner2017}%
  \BibitemOpen
  \bibfield  {author} {\bibinfo {author} {\bibfnamefont {M.}~\bibnamefont
  {Baumgartner}}, \bibinfo {author} {\bibfnamefont {K.}~\bibnamefont
  {Garello}}, \bibinfo {author} {\bibfnamefont {J.}~\bibnamefont {Mendil}},
  \bibinfo {author} {\bibfnamefont {C.~O.}\ \bibnamefont {Avci}}, \bibinfo
  {author} {\bibfnamefont {E.}~\bibnamefont {Grimaldi}}, \bibinfo {author}
  {\bibfnamefont {C.}~\bibnamefont {Murer}}, \bibinfo {author} {\bibfnamefont
  {J.}~\bibnamefont {Feng}}, \bibinfo {author} {\bibfnamefont {M.}~\bibnamefont
  {Gabureac}}, \bibinfo {author} {\bibfnamefont {C.}~\bibnamefont {Stamm}},
  \bibinfo {author} {\bibfnamefont {Y.}~\bibnamefont {Acremann}}, \bibinfo
  {author} {\bibfnamefont {S.}~\bibnamefont {Finizio}}, \bibinfo {author}
  {\bibfnamefont {S.}~\bibnamefont {Wintz}}, \bibinfo {author} {\bibfnamefont
  {J.}~\bibnamefont {Raabe}}, \ and\ \bibinfo {author} {\bibfnamefont
  {P.}~\bibnamefont {Gambardella}},\ }\href
  {http://dx.doi.org/10.1038/nnano.2017.151} {\bibfield  {journal} {\bibinfo
  {journal} {Nat. Nanotechn.}\ }\textbf {\bibinfo {volume} {12}},\ \bibinfo
  {pages} {980} (\bibinfo {year} {2017})}\BibitemShut {NoStop}%
\bibitem [{\citenamefont {Edelstein}(1990)}]{Edelstein1990}%
  \BibitemOpen
  \bibfield  {author} {\bibinfo {author} {\bibfnamefont {V.~M.}\ \bibnamefont
  {Edelstein}},\ }\href {\doibase 10.1016/0038-1098(90)90963-C} {\bibfield
  {journal} {\bibinfo  {journal} {Solid State Commun.}\ }\textbf {\bibinfo
  {volume} {73}},\ \bibinfo {pages} {233} (\bibinfo {year} {1990})}\BibitemShut
  {NoStop}%
\bibitem [{\citenamefont {Bychkov}\ and\ \citenamefont
  {Rashba}(1984)}]{Bychkov1984}%
  \BibitemOpen
  \bibfield  {author} {\bibinfo {author} {\bibfnamefont {Y.~A.}\ \bibnamefont
  {Bychkov}}\ and\ \bibinfo {author} {\bibfnamefont {E.~I.}\ \bibnamefont
  {Rashba}},\ }\href@noop {} {\bibfield  {journal} {\bibinfo  {journal} {JETP
  Lett.}\ }\textbf {\bibinfo {volume} {39}},\ \bibinfo {pages} {78} (\bibinfo
  {year} {1984})}\BibitemShut {NoStop}%
\bibitem [{\citenamefont {Kato}\ \emph
  {et~al.}(2004{\natexlab{b}})\citenamefont {Kato}, \citenamefont {Myers},
  \citenamefont {Gossard},\ and\ \citenamefont {Awschalom}}]{Kato2004}%
  \BibitemOpen
  \bibfield  {author} {\bibinfo {author} {\bibfnamefont {Y.~K.}\ \bibnamefont
  {Kato}}, \bibinfo {author} {\bibfnamefont {R.~C.}\ \bibnamefont {Myers}},
  \bibinfo {author} {\bibfnamefont {A.~C.}\ \bibnamefont {Gossard}}, \ and\
  \bibinfo {author} {\bibfnamefont {D.~D.}\ \bibnamefont {Awschalom}},\ }\href
  {\doibase 10.1103/PhysRevLett.93.176601} {\bibfield  {journal} {\bibinfo
  {journal} {Phys. Rev. Lett.}\ }\textbf {\bibinfo {volume} {93}},\ \bibinfo
  {pages} {176601} (\bibinfo {year} {2004}{\natexlab{b}})}\BibitemShut
  {NoStop}%
\bibitem [{\citenamefont {Silov}\ \emph {et~al.}(2004)\citenamefont {Silov},
  \citenamefont {Blajnov}, \citenamefont {Wolter}, \citenamefont {Hey},
  \citenamefont {Ploog},\ and\ \citenamefont {Averkiev}}]{Silov2004}%
  \BibitemOpen
  \bibfield  {author} {\bibinfo {author} {\bibfnamefont {A.~Y.}\ \bibnamefont
  {Silov}}, \bibinfo {author} {\bibfnamefont {P.~A.}\ \bibnamefont {Blajnov}},
  \bibinfo {author} {\bibfnamefont {J.~H.}\ \bibnamefont {Wolter}}, \bibinfo
  {author} {\bibfnamefont {R.}~\bibnamefont {Hey}}, \bibinfo {author}
  {\bibfnamefont {K.~H.}\ \bibnamefont {Ploog}}, \ and\ \bibinfo {author}
  {\bibfnamefont {N.~S.}\ \bibnamefont {Averkiev}},\ }\href {\doibase
  10.1063/1.1833565} {\bibfield  {journal} {\bibinfo  {journal} {Appl. Phys.
  Lett.}\ }\textbf {\bibinfo {volume} {85}},\ \bibinfo {pages} {5929} (\bibinfo
  {year} {2004})}\BibitemShut {NoStop}%
\bibitem [{\citenamefont {Ganichev}\ \emph {et~al.}(2006)\citenamefont
  {Ganichev}, \citenamefont {Danilov}, \citenamefont {Schneider}, \citenamefont
  {Bel'kov}, \citenamefont {Golub}, \citenamefont {Wegscheider}, \citenamefont
  {Weiss},\ and\ \citenamefont {Prettl}}]{Ganichev2006}%
  \BibitemOpen
  \bibfield  {author} {\bibinfo {author} {\bibfnamefont {S.~D.}\ \bibnamefont
  {Ganichev}}, \bibinfo {author} {\bibfnamefont {S.~N.}\ \bibnamefont
  {Danilov}}, \bibinfo {author} {\bibfnamefont {P.}~\bibnamefont {Schneider}},
  \bibinfo {author} {\bibfnamefont {V.~V.}\ \bibnamefont {Bel'kov}}, \bibinfo
  {author} {\bibfnamefont {L.~E.}\ \bibnamefont {Golub}}, \bibinfo {author}
  {\bibfnamefont {W.}~\bibnamefont {Wegscheider}}, \bibinfo {author}
  {\bibfnamefont {D.}~\bibnamefont {Weiss}}, \ and\ \bibinfo {author}
  {\bibfnamefont {W.}~\bibnamefont {Prettl}},\ }\href@noop {} {\bibfield
  {journal} {\bibinfo  {journal} {J. Magn. Magn. Mater.}\ }\textbf {\bibinfo
  {volume} {300}},\ \bibinfo {pages} {127} (\bibinfo {year}
  {2006})}\BibitemShut {NoStop}%
\bibitem [{\citenamefont {Wadley}\ \emph {et~al.}(2016)\citenamefont {Wadley},
  \citenamefont {Howells}, \citenamefont {\v{Z}elezn{\'y}}, \citenamefont
  {Andrews}, \citenamefont {Hills}, \citenamefont {Campion}, \citenamefont
  {Nov{\'a}k}, \citenamefont {Freimuth}, \citenamefont {Mokrousov},
  \citenamefont {Rushforth}, \citenamefont {Edmonds}, \citenamefont
  {Gallagher},\ and\ \citenamefont {Jungwirth}}]{Wadley2016}%
  \BibitemOpen
  \bibfield  {author} {\bibinfo {author} {\bibfnamefont {P.}~\bibnamefont
  {Wadley}}, \bibinfo {author} {\bibfnamefont {B.}~\bibnamefont {Howells}},
  \bibinfo {author} {\bibfnamefont {J.}~\bibnamefont {\v{Z}elezn{\'y}}},
  \bibinfo {author} {\bibfnamefont {C.}~\bibnamefont {Andrews}}, \bibinfo
  {author} {\bibfnamefont {V.}~\bibnamefont {Hills}}, \bibinfo {author}
  {\bibfnamefont {R.~P.}\ \bibnamefont {Campion}}, \bibinfo {author}
  {\bibfnamefont {V.}~\bibnamefont {Nov{\'a}k}}, \bibinfo {author}
  {\bibfnamefont {F.}~\bibnamefont {Freimuth}}, \bibinfo {author}
  {\bibfnamefont {Y.}~\bibnamefont {Mokrousov}}, \bibinfo {author}
  {\bibfnamefont {A.~W.}\ \bibnamefont {Rushforth}}, \bibinfo {author}
  {\bibfnamefont {K.~W.}\ \bibnamefont {Edmonds}}, \bibinfo {author}
  {\bibfnamefont {B.~L.}\ \bibnamefont {Gallagher}}, \ and\ \bibinfo {author}
  {\bibfnamefont {T.}~\bibnamefont {Jungwirth}},\ }\href {\doibase
  10.1126/science.aab1031} {\bibfield  {journal} {\bibinfo  {journal}
  {Science}\ }\textbf {\bibinfo {volume} {351}},\ \bibinfo {pages} {587}
  (\bibinfo {year} {2016})}\BibitemShut {NoStop}%
\bibitem [{\citenamefont {Olejn{\'{i}}k}\ \emph {et~al.}(2017)\citenamefont
  {Olejn{\'{i}}k}, \citenamefont {Schuler}, \citenamefont {Marti},
  \citenamefont {Nov{\'{a}}k}, \citenamefont {Ka{\v{s}}par}, \citenamefont
  {Wadley}, \citenamefont {Campion}, \citenamefont {Edmonds}, \citenamefont
  {Gallagher}, \citenamefont {Garces}, \citenamefont {Baumgartner},
  \citenamefont {Gambardella},\ and\ \citenamefont {Jungwirth}}]{Olejnik2017}%
  \BibitemOpen
  \bibfield  {author} {\bibinfo {author} {\bibfnamefont {K.}~\bibnamefont
  {Olejn{\'{i}}k}}, \bibinfo {author} {\bibfnamefont {V.}~\bibnamefont
  {Schuler}}, \bibinfo {author} {\bibfnamefont {X.}~\bibnamefont {Marti}},
  \bibinfo {author} {\bibfnamefont {V.}~\bibnamefont {Nov{\'{a}}k}}, \bibinfo
  {author} {\bibfnamefont {Z.}~\bibnamefont {Ka{\v{s}}par}}, \bibinfo {author}
  {\bibfnamefont {P.}~\bibnamefont {Wadley}}, \bibinfo {author} {\bibfnamefont
  {R.~P.}\ \bibnamefont {Campion}}, \bibinfo {author} {\bibfnamefont {K.~W.}\
  \bibnamefont {Edmonds}}, \bibinfo {author} {\bibfnamefont {B.~L.}\
  \bibnamefont {Gallagher}}, \bibinfo {author} {\bibfnamefont {J.}~\bibnamefont
  {Garces}}, \bibinfo {author} {\bibfnamefont {M.}~\bibnamefont {Baumgartner}},
  \bibinfo {author} {\bibfnamefont {P.}~\bibnamefont {Gambardella}}, \ and\
  \bibinfo {author} {\bibfnamefont {T.}~\bibnamefont {Jungwirth}},\ }\href
  {\doibase 10.1038/ncomms15434} {\bibfield  {journal} {\bibinfo  {journal}
  {Nat. Commun.}\ }\textbf {\bibinfo {volume} {8}},\ \bibinfo {pages} {15434}
  (\bibinfo {year} {2017})}\BibitemShut {NoStop}%
\bibitem [{\citenamefont {Bodnar}\ \emph {et~al.}(2018)\citenamefont {Bodnar},
  \citenamefont {{\v{S}}mejkal}, \citenamefont {Turek}, \citenamefont
  {Jungwirth}, \citenamefont {Gomonay}, \citenamefont {Sinova}, \citenamefont
  {Sapozhnik}, \citenamefont {Elmers}, \citenamefont {Kl{\"{a}}ui},\ and\
  \citenamefont {Jourdan}}]{Bodnar2018}%
  \BibitemOpen
  \bibfield  {author} {\bibinfo {author} {\bibfnamefont {S.~Y.}\ \bibnamefont
  {Bodnar}}, \bibinfo {author} {\bibfnamefont {L.}~\bibnamefont
  {{\v{S}}mejkal}}, \bibinfo {author} {\bibfnamefont {I.}~\bibnamefont
  {Turek}}, \bibinfo {author} {\bibfnamefont {T.}~\bibnamefont {Jungwirth}},
  \bibinfo {author} {\bibfnamefont {O.}~\bibnamefont {Gomonay}}, \bibinfo
  {author} {\bibfnamefont {J.}~\bibnamefont {Sinova}}, \bibinfo {author}
  {\bibfnamefont {A.~A.}\ \bibnamefont {Sapozhnik}}, \bibinfo {author}
  {\bibfnamefont {H.-J.}\ \bibnamefont {Elmers}}, \bibinfo {author}
  {\bibfnamefont {M.}~\bibnamefont {Kl{\"{a}}ui}}, \ and\ \bibinfo {author}
  {\bibfnamefont {M.}~\bibnamefont {Jourdan}},\ }\href {\doibase
  10.1038/s41467-017-02780-x} {\bibfield  {journal} {\bibinfo  {journal} {Nat.
  Commun.}\ }\textbf {\bibinfo {volume} {9}},\ \bibinfo {pages} {348} (\bibinfo
  {year} {2018})}\BibitemShut {NoStop}%
\bibitem [{\citenamefont {Meinert}\ \emph {et~al.}(2018)\citenamefont
  {Meinert}, \citenamefont {Graulich},\ and\ \citenamefont
  {Matalla-Wagner}}]{Meinert2018}%
  \BibitemOpen
  \bibfield  {author} {\bibinfo {author} {\bibfnamefont {M.}~\bibnamefont
  {Meinert}}, \bibinfo {author} {\bibfnamefont {D.}~\bibnamefont {Graulich}}, \
  and\ \bibinfo {author} {\bibfnamefont {T.}~\bibnamefont {Matalla-Wagner}},\
  }\href {\doibase 10.1103/PhysRevApplied.9.064040} {\bibfield  {journal}
  {\bibinfo  {journal} {Phys. Rev. Applied}\ }\textbf {\bibinfo {volume} {9}},\
  \bibinfo {pages} {064040} (\bibinfo {year} {2018})}\BibitemShut {NoStop}%
\bibitem [{\citenamefont {Godinho}\ \emph {et~al.}(2018)\citenamefont
  {Godinho}, \citenamefont {Reichlov{\'{a}}}, \citenamefont {Kriegner},
  \citenamefont {Nov{\'{a}}k}, \citenamefont {Olejn{\'{i}}k}, \citenamefont
  {Ka{\v{s}}par}, \citenamefont {{\v{S}}ob{\'{a}}?}, \citenamefont {Wadley},
  \citenamefont {Campion}, \citenamefont {Otxoa}, \citenamefont {Roy},
  \citenamefont {{\v{Z}}elezn{\'{y}}}, \citenamefont {Jungwirth},\ and\
  \citenamefont {Wunderlich}}]{Godinho2018}%
  \BibitemOpen
  \bibfield  {author} {\bibinfo {author} {\bibfnamefont {J.}~\bibnamefont
  {Godinho}}, \bibinfo {author} {\bibfnamefont {H.}~\bibnamefont
  {Reichlov{\'{a}}}}, \bibinfo {author} {\bibfnamefont {D.}~\bibnamefont
  {Kriegner}}, \bibinfo {author} {\bibfnamefont {V.}~\bibnamefont
  {Nov{\'{a}}k}}, \bibinfo {author} {\bibfnamefont {K.}~\bibnamefont
  {Olejn{\'{i}}k}}, \bibinfo {author} {\bibfnamefont {Z.}~\bibnamefont
  {Ka{\v{s}}par}}, \bibinfo {author} {\bibfnamefont {Z.}~\bibnamefont
  {{\v{S}}ob{\'{a}}?}}, \bibinfo {author} {\bibfnamefont {P.}~\bibnamefont
  {Wadley}}, \bibinfo {author} {\bibfnamefont {R.~P.}\ \bibnamefont {Campion}},
  \bibinfo {author} {\bibfnamefont {R.~M.}\ \bibnamefont {Otxoa}}, \bibinfo
  {author} {\bibfnamefont {P.~E.}\ \bibnamefont {Roy}}, \bibinfo {author}
  {\bibfnamefont {J.}~\bibnamefont {{\v{Z}}elezn{\'{y}}}}, \bibinfo {author}
  {\bibfnamefont {T.}~\bibnamefont {Jungwirth}}, \ and\ \bibinfo {author}
  {\bibfnamefont {J.}~\bibnamefont {Wunderlich}},\ }\href {\doibase
  10.1038/s41467-018-07092-2} {\bibfield  {journal} {\bibinfo  {journal} {Nat.
  Commun.}\ }\textbf {\bibinfo {volume} {9}},\ \bibinfo {pages} {4686}
  (\bibinfo {year} {2018})}\BibitemShut {NoStop}%
\bibitem [{\citenamefont {Grzybowski}\ \emph {et~al.}(2017)\citenamefont
  {Grzybowski}, \citenamefont {Wadley}, \citenamefont {Edmonds}, \citenamefont
  {Beardsley}, \citenamefont {Hills}, \citenamefont {Campion}, \citenamefont
  {Gallagher}, \citenamefont {Chauhan}, \citenamefont {Novak}, \citenamefont
  {Jungwirth}, \citenamefont {Maccherozzi},\ and\ \citenamefont
  {Dhesi}}]{Grzybowski2017}%
  \BibitemOpen
  \bibfield  {author} {\bibinfo {author} {\bibfnamefont {M.~J.}\ \bibnamefont
  {Grzybowski}}, \bibinfo {author} {\bibfnamefont {P.}~\bibnamefont {Wadley}},
  \bibinfo {author} {\bibfnamefont {K.~W.}\ \bibnamefont {Edmonds}}, \bibinfo
  {author} {\bibfnamefont {R.}~\bibnamefont {Beardsley}}, \bibinfo {author}
  {\bibfnamefont {V.}~\bibnamefont {Hills}}, \bibinfo {author} {\bibfnamefont
  {R.~P.}\ \bibnamefont {Campion}}, \bibinfo {author} {\bibfnamefont {B.~L.}\
  \bibnamefont {Gallagher}}, \bibinfo {author} {\bibfnamefont {J.~S.}\
  \bibnamefont {Chauhan}}, \bibinfo {author} {\bibfnamefont {V.}~\bibnamefont
  {Novak}}, \bibinfo {author} {\bibfnamefont {T.}~\bibnamefont {Jungwirth}},
  \bibinfo {author} {\bibfnamefont {F.}~\bibnamefont {Maccherozzi}}, \ and\
  \bibinfo {author} {\bibfnamefont {S.~S.}\ \bibnamefont {Dhesi}},\ }\href
  {https://journals.aps.org/prl/pdf/10.1103/PhysRevLett.118.057701} {\bibfield
  {journal} {\bibinfo  {journal} {Phys. Rev. Lett.}\ }\textbf {\bibinfo
  {volume} {118}} (\bibinfo {year} {2017})}\BibitemShut {NoStop}%
\bibitem [{\citenamefont {Wadley}\ \emph {et~al.}(2018)\citenamefont {Wadley},
  \citenamefont {Reimers}, \citenamefont {Grzybowski}, \citenamefont {Andrews},
  \citenamefont {Wang}, \citenamefont {Chauhan}, \citenamefont {Gallagher},
  \citenamefont {Campion}, \citenamefont {Edmonds}, \citenamefont {Dhesi},
  \citenamefont {Maccherozzi}, \citenamefont {Nov{\'a}k}, \citenamefont
  {Wunderlich},\ and\ \citenamefont {Jungwirth}}]{Wadley2018}%
  \BibitemOpen
  \bibfield  {author} {\bibinfo {author} {\bibfnamefont {P.}~\bibnamefont
  {Wadley}}, \bibinfo {author} {\bibfnamefont {S.}~\bibnamefont {Reimers}},
  \bibinfo {author} {\bibfnamefont {M.~J.}\ \bibnamefont {Grzybowski}},
  \bibinfo {author} {\bibfnamefont {C.}~\bibnamefont {Andrews}}, \bibinfo
  {author} {\bibfnamefont {M.}~\bibnamefont {Wang}}, \bibinfo {author}
  {\bibfnamefont {J.~S.}\ \bibnamefont {Chauhan}}, \bibinfo {author}
  {\bibfnamefont {B.~L.}\ \bibnamefont {Gallagher}}, \bibinfo {author}
  {\bibfnamefont {R.~P.}\ \bibnamefont {Campion}}, \bibinfo {author}
  {\bibfnamefont {K.~W.}\ \bibnamefont {Edmonds}}, \bibinfo {author}
  {\bibfnamefont {S.~S.}\ \bibnamefont {Dhesi}}, \bibinfo {author}
  {\bibfnamefont {F.}~\bibnamefont {Maccherozzi}}, \bibinfo {author}
  {\bibfnamefont {V.}~\bibnamefont {Nov{\'a}k}}, \bibinfo {author}
  {\bibfnamefont {J.}~\bibnamefont {Wunderlich}}, \ and\ \bibinfo {author}
  {\bibfnamefont {T.}~\bibnamefont {Jungwirth}},\ }\href@noop {} {\bibfield
  {journal} {\bibinfo  {journal} {Nat. Nanotechn.}\ }\textbf {\bibinfo {volume}
  {13}},\ \bibinfo {pages} {362} (\bibinfo {year} {2018})}\BibitemShut
  {NoStop}%
\bibitem [{\citenamefont {{\v{Z}}elezn{\'{y}}}\ \emph
  {et~al.}(2014)\citenamefont {{\v{Z}}elezn{\'{y}}}, \citenamefont {Gao},
  \citenamefont {V{\'{y}}born{\'{y}}}, \citenamefont {Zemen}, \citenamefont
  {Ma{\v{s}}ek}, \citenamefont {Manchon}, \citenamefont {Wunderlich},
  \citenamefont {Sinova},\ and\ \citenamefont {Jungwirth}}]{Zelezny2014}%
  \BibitemOpen
  \bibfield  {author} {\bibinfo {author} {\bibfnamefont {J.}~\bibnamefont
  {{\v{Z}}elezn{\'{y}}}}, \bibinfo {author} {\bibfnamefont {H.}~\bibnamefont
  {Gao}}, \bibinfo {author} {\bibfnamefont {K.}~\bibnamefont
  {V{\'{y}}born{\'{y}}}}, \bibinfo {author} {\bibfnamefont {J.}~\bibnamefont
  {Zemen}}, \bibinfo {author} {\bibfnamefont {J.}~\bibnamefont {Ma{\v{s}}ek}},
  \bibinfo {author} {\bibfnamefont {A.}~\bibnamefont {Manchon}}, \bibinfo
  {author} {\bibfnamefont {J.}~\bibnamefont {Wunderlich}}, \bibinfo {author}
  {\bibfnamefont {J.}~\bibnamefont {Sinova}}, \ and\ \bibinfo {author}
  {\bibfnamefont {T.}~\bibnamefont {Jungwirth}},\ }\href {\doibase
  10.1103/PhysRevLett.113.157201} {\bibfield  {journal} {\bibinfo  {journal}
  {Phys. Rev. Lett.}\ }\textbf {\bibinfo {volume} {113}},\ \bibinfo {pages}
  {157201} (\bibinfo {year} {2014})}\BibitemShut {NoStop}%
\bibitem [{\citenamefont {{\v{Z}}elezn{\'{y}}}\ \emph
  {et~al.}(2017)\citenamefont {{\v{Z}}elezn{\'{y}}}, \citenamefont {Gao},
  \citenamefont {Manchon}, \citenamefont {Freimuth}, \citenamefont {Mokrousov},
  \citenamefont {Zemen}, \citenamefont {Ma{\v{s}}ek}, \citenamefont {Sinova},\
  and\ \citenamefont {Jungwirth}}]{Zelezny2017}%
  \BibitemOpen
  \bibfield  {author} {\bibinfo {author} {\bibfnamefont {J.}~\bibnamefont
  {{\v{Z}}elezn{\'{y}}}}, \bibinfo {author} {\bibfnamefont {H.}~\bibnamefont
  {Gao}}, \bibinfo {author} {\bibfnamefont {A.}~\bibnamefont {Manchon}},
  \bibinfo {author} {\bibfnamefont {F.}~\bibnamefont {Freimuth}}, \bibinfo
  {author} {\bibfnamefont {Y.}~\bibnamefont {Mokrousov}}, \bibinfo {author}
  {\bibfnamefont {J.}~\bibnamefont {Zemen}}, \bibinfo {author} {\bibfnamefont
  {J.}~\bibnamefont {Ma{\v{s}}ek}}, \bibinfo {author} {\bibfnamefont
  {J.}~\bibnamefont {Sinova}}, \ and\ \bibinfo {author} {\bibfnamefont
  {T.}~\bibnamefont {Jungwirth}},\ }\href {\doibase 10.1103/PhysRevB.95.014403}
  {\bibfield  {journal} {\bibinfo  {journal} {Phys. Rev. B}\ }\textbf {\bibinfo
  {volume} {95}},\ \bibinfo {pages} {014403} (\bibinfo {year}
  {2017})}\BibitemShut {NoStop}%
\bibitem [{\citenamefont {Johansson}\ \emph {et~al.}(2018)\citenamefont
  {Johansson}, \citenamefont {Henk},\ and\ \citenamefont
  {Mertig}}]{Johansson2018}%
  \BibitemOpen
  \bibfield  {author} {\bibinfo {author} {\bibfnamefont {A.}~\bibnamefont
  {Johansson}}, \bibinfo {author} {\bibfnamefont {J.}~\bibnamefont {Henk}}, \
  and\ \bibinfo {author} {\bibfnamefont {I.}~\bibnamefont {Mertig}},\ }\href
  {\doibase 10.1103/PhysRevB.97.085417} {\bibfield  {journal} {\bibinfo
  {journal} {Phys. Rev. B}\ }\textbf {\bibinfo {volume} {97}},\ \bibinfo
  {pages} {085417} (\bibinfo {year} {2018})}\BibitemShut {NoStop}%
\bibitem [{\citenamefont {Blaha}\ \emph {et~al.}(2018)\citenamefont {Blaha},
  \citenamefont {Schwarz}, \citenamefont {Madsen}, \citenamefont {Kvasnicka},
  \citenamefont {Luitz}, \citenamefont {Laskowski}, \citenamefont {Tran},\ and\
  \citenamefont {Marks}}]{Blaha2018}%
  \BibitemOpen
  \bibfield  {author} {\bibinfo {author} {\bibfnamefont {P.}~\bibnamefont
  {Blaha}}, \bibinfo {author} {\bibfnamefont {K.}~\bibnamefont {Schwarz}},
  \bibinfo {author} {\bibfnamefont {G.~K.~H.}\ \bibnamefont {Madsen}}, \bibinfo
  {author} {\bibfnamefont {D.}~\bibnamefont {Kvasnicka}}, \bibinfo {author}
  {\bibfnamefont {J.}~\bibnamefont {Luitz}}, \bibinfo {author} {\bibfnamefont
  {R.}~\bibnamefont {Laskowski}}, \bibinfo {author} {\bibfnamefont
  {F.}~\bibnamefont {Tran}}, \ and\ \bibinfo {author} {\bibfnamefont {L.~D.}\
  \bibnamefont {Marks}},\ }\href {\doibase citeulike-article-id:6205108} {\emph
  {\bibinfo {title} {{WIEN2k, An Augmented Plane Wave + Local Orbitals Program
  for Calculating Crystal Properties \textrm{(Techn. Universit{\"{a}}t Wien,
  Austria)}}}}},\ \bibinfo {number} {ISBN 3-9501031-1-2}\ (\bibinfo {year}
  {2018})\BibitemShut {NoStop}%
\bibitem [{\citenamefont {Olejn{\'{i}}k}\ \emph {et~al.}(2018)\citenamefont
  {Olejn{\'{i}}k}, \citenamefont {Seifert}, \citenamefont {Ka{\v{s}}par},
  \citenamefont {Nov{\'{a}}k}, \citenamefont {Wadley}, \citenamefont {Campion},
  \citenamefont {Baumgartner}, \citenamefont {Gambardella}, \citenamefont
  {N{\v{e}}mec}, \citenamefont {Wunderlich}, \citenamefont {Sinova},
  \citenamefont {Ku{\v{z}}el}, \citenamefont {M{\"{u}}ller}, \citenamefont
  {Kampfrath},\ and\ \citenamefont {Jungwirth}}]{Olejnik2018}%
  \BibitemOpen
  \bibfield  {author} {\bibinfo {author} {\bibfnamefont {K.}~\bibnamefont
  {Olejn{\'{i}}k}}, \bibinfo {author} {\bibfnamefont {T.}~\bibnamefont
  {Seifert}}, \bibinfo {author} {\bibfnamefont {Z.}~\bibnamefont
  {Ka{\v{s}}par}}, \bibinfo {author} {\bibfnamefont {V.}~\bibnamefont
  {Nov{\'{a}}k}}, \bibinfo {author} {\bibfnamefont {P.}~\bibnamefont {Wadley}},
  \bibinfo {author} {\bibfnamefont {R.~P.}\ \bibnamefont {Campion}}, \bibinfo
  {author} {\bibfnamefont {M.}~\bibnamefont {Baumgartner}}, \bibinfo {author}
  {\bibfnamefont {P.}~\bibnamefont {Gambardella}}, \bibinfo {author}
  {\bibfnamefont {P.}~\bibnamefont {N{\v{e}}mec}}, \bibinfo {author}
  {\bibfnamefont {J.}~\bibnamefont {Wunderlich}}, \bibinfo {author}
  {\bibfnamefont {J.}~\bibnamefont {Sinova}}, \bibinfo {author} {\bibfnamefont
  {P.}~\bibnamefont {Ku{\v{z}}el}}, \bibinfo {author} {\bibfnamefont
  {M.}~\bibnamefont {M{\"{u}}ller}}, \bibinfo {author} {\bibfnamefont
  {T.}~\bibnamefont {Kampfrath}}, \ and\ \bibinfo {author} {\bibfnamefont
  {T.}~\bibnamefont {Jungwirth}},\ }\href {\doibase 10.1126/sciadv.aar3566}
  {\bibfield  {journal} {\bibinfo  {journal} {Sci. Adv.}\ }\textbf {\bibinfo
  {volume} {4}},\ \bibinfo {pages} {eaar3566} (\bibinfo {year}
  {2018})}\BibitemShut {NoStop}%
\bibitem [{\citenamefont {Wadley}\ \emph {et~al.}(2013)\citenamefont {Wadley},
  \citenamefont {Nov{\'a}k}, \citenamefont {Campion}, \citenamefont {Rinaldi},
  \citenamefont {Mart{\'i}}, \citenamefont {Reichlov{\'a}}, \citenamefont
  {\v{Z}elezn{\'y}}, \citenamefont {Gazquez}, \citenamefont {Roldan},
  \citenamefont {Varela}, \citenamefont {Khalyavin}, \citenamefont {Langridge},
  \citenamefont {Kriegner}, \citenamefont {M{\'a}ca}, \citenamefont
  {Ma\v{s}ek}, \citenamefont {Bertacco}, \citenamefont {Hol{\'y}},
  \citenamefont {Rushforth}, \citenamefont {Edmonds}, \citenamefont
  {Gallagher}, \citenamefont {Foxon}, \citenamefont {Wunderlich},\ and\
  \citenamefont {Jungwirth}}]{Wadley2013}%
  \BibitemOpen
  \bibfield  {author} {\bibinfo {author} {\bibfnamefont {P.}~\bibnamefont
  {Wadley}}, \bibinfo {author} {\bibfnamefont {V.}~\bibnamefont {Nov{\'a}k}},
  \bibinfo {author} {\bibfnamefont {R.~P.}\ \bibnamefont {Campion}}, \bibinfo
  {author} {\bibfnamefont {C.}~\bibnamefont {Rinaldi}}, \bibinfo {author}
  {\bibfnamefont {X.}~\bibnamefont {Mart{\'i}}}, \bibinfo {author}
  {\bibfnamefont {H.}~\bibnamefont {Reichlov{\'a}}}, \bibinfo {author}
  {\bibfnamefont {J.}~\bibnamefont {\v{Z}elezn{\'y}}}, \bibinfo {author}
  {\bibfnamefont {J.}~\bibnamefont {Gazquez}}, \bibinfo {author} {\bibfnamefont
  {M.~A.}\ \bibnamefont {Roldan}}, \bibinfo {author} {\bibfnamefont
  {M.}~\bibnamefont {Varela}}, \bibinfo {author} {\bibfnamefont
  {D.}~\bibnamefont {Khalyavin}}, \bibinfo {author} {\bibfnamefont
  {S.}~\bibnamefont {Langridge}}, \bibinfo {author} {\bibfnamefont
  {D.}~\bibnamefont {Kriegner}}, \bibinfo {author} {\bibfnamefont
  {F.}~\bibnamefont {M{\'a}ca}}, \bibinfo {author} {\bibfnamefont
  {J.}~\bibnamefont {Ma\v{s}ek}}, \bibinfo {author} {\bibfnamefont
  {R.}~\bibnamefont {Bertacco}}, \bibinfo {author} {\bibfnamefont
  {V.}~\bibnamefont {Hol{\'y}}}, \bibinfo {author} {\bibfnamefont {A.~W.}\
  \bibnamefont {Rushforth}}, \bibinfo {author} {\bibfnamefont {K.~W.}\
  \bibnamefont {Edmonds}}, \bibinfo {author} {\bibfnamefont {B.~L.}\
  \bibnamefont {Gallagher}}, \bibinfo {author} {\bibfnamefont {C.}~\bibnamefont
  {Foxon}}, \bibinfo {author} {\bibfnamefont {J.}~\bibnamefont {Wunderlich}}, \
  and\ \bibinfo {author} {\bibfnamefont {T.}~\bibnamefont {Jungwirth}},\
  }\href@noop {} {\bibfield  {journal} {\bibinfo  {journal} {Nat. Commun.}\
  }\textbf {\bibinfo {volume} {4}},\ \bibinfo {pages} {2322} (\bibinfo {year}
  {2013})}\BibitemShut {NoStop}%
\bibitem [{\citenamefont {Wadley}\ \emph {et~al.}(2015)\citenamefont {Wadley},
  \citenamefont {Hills}, \citenamefont {Shahedkhah}, \citenamefont {Edmonds},
  \citenamefont {Campion}, \citenamefont {Nov{\'{a}}k}, \citenamefont
  {Ouladdiaf}, \citenamefont {Khalyavin}, \citenamefont {Langridge},
  \citenamefont {Saidl}, \citenamefont {N\v{e}mec}, \citenamefont {Rushforth},
  \citenamefont {Gallagher}, \citenamefont {Dhesi}, \citenamefont
  {Maccherozzi}, \citenamefont {{\v{Z}}elezn{\'{y}}},\ and\ \citenamefont
  {Jungwirth}}]{Wadley2015}%
  \BibitemOpen
  \bibfield  {author} {\bibinfo {author} {\bibfnamefont {P.}~\bibnamefont
  {Wadley}}, \bibinfo {author} {\bibfnamefont {V.}~\bibnamefont {Hills}},
  \bibinfo {author} {\bibfnamefont {M.~R.}\ \bibnamefont {Shahedkhah}},
  \bibinfo {author} {\bibfnamefont {K.~W.}\ \bibnamefont {Edmonds}}, \bibinfo
  {author} {\bibfnamefont {R.~P.}\ \bibnamefont {Campion}}, \bibinfo {author}
  {\bibfnamefont {V.}~\bibnamefont {Nov{\'{a}}k}}, \bibinfo {author}
  {\bibfnamefont {B.}~\bibnamefont {Ouladdiaf}}, \bibinfo {author}
  {\bibfnamefont {D.}~\bibnamefont {Khalyavin}}, \bibinfo {author}
  {\bibfnamefont {S.}~\bibnamefont {Langridge}}, \bibinfo {author}
  {\bibfnamefont {V.}~\bibnamefont {Saidl}}, \bibinfo {author} {\bibfnamefont
  {P.}~\bibnamefont {N\v{e}mec}}, \bibinfo {author} {\bibfnamefont {A.~W.}\
  \bibnamefont {Rushforth}}, \bibinfo {author} {\bibfnamefont {B.~L.}\
  \bibnamefont {Gallagher}}, \bibinfo {author} {\bibfnamefont {S.~S.}\
  \bibnamefont {Dhesi}}, \bibinfo {author} {\bibfnamefont {F.}~\bibnamefont
  {Maccherozzi}}, \bibinfo {author} {\bibfnamefont {J.}~\bibnamefont
  {{\v{Z}}elezn{\'{y}}}}, \ and\ \bibinfo {author} {\bibfnamefont
  {T.}~\bibnamefont {Jungwirth}},\ }\href {\doibase 10.1038/srep17079}
  {\bibfield  {journal} {\bibinfo  {journal} {Sci. Rep.}\ }\textbf {\bibinfo
  {volume} {5}},\ \bibinfo {pages} {17079} (\bibinfo {year}
  {2015})}\BibitemShut {NoStop}%
\bibitem [{\citenamefont {Manchon}\ \emph {et~al.}(2015)\citenamefont
  {Manchon}, \citenamefont {Koo}, \citenamefont {Nitta}, \citenamefont
  {Frolov},\ and\ \citenamefont {Duine}}]{Manchon2015}%
  \BibitemOpen
  \bibfield  {author} {\bibinfo {author} {\bibfnamefont {A.}~\bibnamefont
  {Manchon}}, \bibinfo {author} {\bibfnamefont {H.~C.}\ \bibnamefont {Koo}},
  \bibinfo {author} {\bibfnamefont {J.}~\bibnamefont {Nitta}}, \bibinfo
  {author} {\bibfnamefont {S.~M.}\ \bibnamefont {Frolov}}, \ and\ \bibinfo
  {author} {\bibfnamefont {R.~A.}\ \bibnamefont {Duine}},\ }\href@noop {}
  {\bibfield  {journal} {\bibinfo  {journal} {Nat. Mater.}\ }\textbf {\bibinfo
  {volume} {14}},\ \bibinfo {pages} {871} (\bibinfo {year} {2015})}\BibitemShut
  {NoStop}%
\bibitem [{\citenamefont {Ciccarelli}\ \emph {et~al.}(2016)\citenamefont
  {Ciccarelli}, \citenamefont {Anderson}, \citenamefont {Tshitoyan},
  \citenamefont {Ferguson}, \citenamefont {Gerhard}, \citenamefont {Gould},
  \citenamefont {Molenkamp}, \citenamefont {Gayles}, \citenamefont
  {{\v{Z}}elezn{\'{y}}}, \citenamefont {{\v{S}}mejkal}, \citenamefont {Yuan},
  \citenamefont {Sinova}, \citenamefont {Freimuth},\ and\ \citenamefont
  {Jungwirth}}]{Ciccarelli2016}%
  \BibitemOpen
  \bibfield  {author} {\bibinfo {author} {\bibfnamefont {C.}~\bibnamefont
  {Ciccarelli}}, \bibinfo {author} {\bibfnamefont {L.}~\bibnamefont
  {Anderson}}, \bibinfo {author} {\bibfnamefont {V.}~\bibnamefont {Tshitoyan}},
  \bibinfo {author} {\bibfnamefont {A.~J.}\ \bibnamefont {Ferguson}}, \bibinfo
  {author} {\bibfnamefont {F.}~\bibnamefont {Gerhard}}, \bibinfo {author}
  {\bibfnamefont {C.}~\bibnamefont {Gould}}, \bibinfo {author} {\bibfnamefont
  {L.~W.}\ \bibnamefont {Molenkamp}}, \bibinfo {author} {\bibfnamefont
  {J.}~\bibnamefont {Gayles}}, \bibinfo {author} {\bibfnamefont
  {J.}~\bibnamefont {{\v{Z}}elezn{\'{y}}}}, \bibinfo {author} {\bibfnamefont
  {L.}~\bibnamefont {{\v{S}}mejkal}}, \bibinfo {author} {\bibfnamefont
  {Z.}~\bibnamefont {Yuan}}, \bibinfo {author} {\bibfnamefont {J.}~\bibnamefont
  {Sinova}}, \bibinfo {author} {\bibfnamefont {F.}~\bibnamefont {Freimuth}}, \
  and\ \bibinfo {author} {\bibfnamefont {T.}~\bibnamefont {Jungwirth}},\ }\href
  {\doibase 10.1038/nphys3772} {\bibfield  {journal} {\bibinfo  {journal} {Nat.
  Phys.}\ }\textbf {\bibinfo {volume} {12}},\ \bibinfo {pages} {855} (\bibinfo
  {year} {2016})}\BibitemShut {NoStop}%
\bibitem [{\citenamefont {Kurebayashi}\ \emph {et~al.}(2014)\citenamefont
  {Kurebayashi}, \citenamefont {Sinova}, \citenamefont {Fang}, \citenamefont
  {Irvine}, \citenamefont {Skinner}, \citenamefont {Wunderlich}, \citenamefont
  {Nov{\'{a}}k}, \citenamefont {Campion}, \citenamefont {Gallagher},
  \citenamefont {Vehstedt}, \citenamefont {Z{\^{a}}rbo}, \citenamefont
  {V{\'{y}}born{\'{y}}}, \citenamefont {Ferguson},\ and\ \citenamefont
  {Jungwirth}}]{Kurebayashi2014}%
  \BibitemOpen
  \bibfield  {author} {\bibinfo {author} {\bibfnamefont {H.}~\bibnamefont
  {Kurebayashi}}, \bibinfo {author} {\bibfnamefont {J.}~\bibnamefont {Sinova}},
  \bibinfo {author} {\bibfnamefont {D.}~\bibnamefont {Fang}}, \bibinfo {author}
  {\bibfnamefont {A.~C.}\ \bibnamefont {Irvine}}, \bibinfo {author}
  {\bibfnamefont {T.~D.}\ \bibnamefont {Skinner}}, \bibinfo {author}
  {\bibfnamefont {J.}~\bibnamefont {Wunderlich}}, \bibinfo {author}
  {\bibfnamefont {V.}~\bibnamefont {Nov{\'{a}}k}}, \bibinfo {author}
  {\bibfnamefont {R.~P.}\ \bibnamefont {Campion}}, \bibinfo {author}
  {\bibfnamefont {B.~L.}\ \bibnamefont {Gallagher}}, \bibinfo {author}
  {\bibfnamefont {E.~K.}\ \bibnamefont {Vehstedt}}, \bibinfo {author}
  {\bibfnamefont {L.~P.}\ \bibnamefont {Z{\^{a}}rbo}}, \bibinfo {author}
  {\bibfnamefont {K.}~\bibnamefont {V{\'{y}}born{\'{y}}}}, \bibinfo {author}
  {\bibfnamefont {A.~J.}\ \bibnamefont {Ferguson}}, \ and\ \bibinfo {author}
  {\bibfnamefont {T.}~\bibnamefont {Jungwirth}},\ }\href {\doibase
  10.1038/nnano.2014.15} {\bibfield  {journal} {\bibinfo  {journal} {Nat.
  Nanotechn.}\ }\textbf {\bibinfo {volume} {9}},\ \bibinfo {pages} {211}
  (\bibinfo {year} {2014})}\BibitemShut {NoStop}%
\bibitem [{\citenamefont {Go}\ \emph {et~al.}(2017)\citenamefont {Go},
  \citenamefont {Hanke}, \citenamefont {Buhl}, \citenamefont {Freimuth},
  \citenamefont {Bihlmayer}, \citenamefont {Lee}, \citenamefont {Mokrousov},\
  and\ \citenamefont {Bl{\"u}gel}}]{Go2017}%
  \BibitemOpen
  \bibfield  {author} {\bibinfo {author} {\bibfnamefont {D.}~\bibnamefont
  {Go}}, \bibinfo {author} {\bibfnamefont {J.-P.}\ \bibnamefont {Hanke}},
  \bibinfo {author} {\bibfnamefont {P.~M.}\ \bibnamefont {Buhl}}, \bibinfo
  {author} {\bibfnamefont {F.}~\bibnamefont {Freimuth}}, \bibinfo {author}
  {\bibfnamefont {G.}~\bibnamefont {Bihlmayer}}, \bibinfo {author}
  {\bibfnamefont {H.-W.}\ \bibnamefont {Lee}}, \bibinfo {author} {\bibfnamefont
  {Y.}~\bibnamefont {Mokrousov}}, \ and\ \bibinfo {author} {\bibfnamefont
  {S.}~\bibnamefont {Bl{\"u}gel}},\ }\href@noop {} {\bibfield  {journal}
  {\bibinfo  {journal} {Sci. Rep.}\ }\textbf {\bibinfo {volume} {7}},\ \bibinfo
  {pages} {46742} (\bibinfo {year} {2017})}\BibitemShut {NoStop}%
\bibitem [{\citenamefont {Hanke}\ \emph {et~al.}(2017)\citenamefont {Hanke},
  \citenamefont {Freimuth}, \citenamefont {Bl{\"u}gel},\ and\ \citenamefont
  {Mokrousov}}]{Hanke2017}%
  \BibitemOpen
  \bibfield  {author} {\bibinfo {author} {\bibfnamefont {J.-P.}\ \bibnamefont
  {Hanke}}, \bibinfo {author} {\bibfnamefont {F.}~\bibnamefont {Freimuth}},
  \bibinfo {author} {\bibfnamefont {S.}~\bibnamefont {Bl{\"u}gel}}, \ and\
  \bibinfo {author} {\bibfnamefont {Y.}~\bibnamefont {Mokrousov}},\ }\href@noop
  {} {\bibfield  {journal} {\bibinfo  {journal} {Sci. Rep.}\ }\textbf {\bibinfo
  {volume} {7}},\ \bibinfo {pages} {41078} (\bibinfo {year}
  {2017})}\BibitemShut {NoStop}%
\bibitem [{\citenamefont {Yoda}\ \emph {et~al.}(2018)\citenamefont {Yoda},
  \citenamefont {Yokoyama},\ and\ \citenamefont {Murakami}}]{Yoda2018}%
  \BibitemOpen
  \bibfield  {author} {\bibinfo {author} {\bibfnamefont {T.}~\bibnamefont
  {Yoda}}, \bibinfo {author} {\bibfnamefont {T.}~\bibnamefont {Yokoyama}}, \
  and\ \bibinfo {author} {\bibfnamefont {S.}~\bibnamefont {Murakami}},\
  }\href@noop {} {\bibfield  {journal} {\bibinfo  {journal} {Nano Lett.}\
  }\textbf {\bibinfo {volume} {18}},\ \bibinfo {pages} {916} (\bibinfo {year}
  {2018})}\BibitemShut {NoStop}%
\bibitem [{\citenamefont {Kittel}(1946)}]{Kittel1946}%
  \BibitemOpen
  \bibfield  {author} {\bibinfo {author} {\bibfnamefont {C.}~\bibnamefont
  {Kittel}},\ }\href {\doibase 10.1103/PhysRev.70.281} {\bibfield  {journal}
  {\bibinfo  {journal} {Phys. Rev.}\ }\textbf {\bibinfo {volume} {70}},\
  \bibinfo {pages} {281} (\bibinfo {year} {1946})}\BibitemShut {NoStop}%
\bibitem [{\citenamefont {Wienholdt}\ \emph {et~al.}(2013)\citenamefont
  {Wienholdt}, \citenamefont {Hinzke}, \citenamefont {Carva}, \citenamefont
  {Oppeneer},\ and\ \citenamefont {Nowak}}]{Wienholdt2013}%
  \BibitemOpen
  \bibfield  {author} {\bibinfo {author} {\bibfnamefont {S.}~\bibnamefont
  {Wienholdt}}, \bibinfo {author} {\bibfnamefont {D.}~\bibnamefont {Hinzke}},
  \bibinfo {author} {\bibfnamefont {K.}~\bibnamefont {Carva}}, \bibinfo
  {author} {\bibfnamefont {P.~M.}\ \bibnamefont {Oppeneer}}, \ and\ \bibinfo
  {author} {\bibfnamefont {U.}~\bibnamefont {Nowak}},\ }\href {\doibase
  10.1103/PhysRevB.88.020406} {\bibfield  {journal} {\bibinfo  {journal} {Phys.
  Rev. B}\ }\textbf {\bibinfo {volume} {88}},\ \bibinfo {pages} {020406}
  (\bibinfo {year} {2013})}\BibitemShut {NoStop}%
\bibitem [{\citenamefont {Li}\ \emph {et~al.}(2018)\citenamefont {Li},
  \citenamefont {Chen}, \citenamefont {Zuo}, \citenamefont {Yun}, \citenamefont
  {Cui}, \citenamefont {Wu}, \citenamefont {Guo}, \citenamefont {Yang},
  \citenamefont {Wang},\ and\ \citenamefont {Xi}}]{Li2018}%
  \BibitemOpen
  \bibfield  {author} {\bibinfo {author} {\bibfnamefont {D.}~\bibnamefont
  {Li}}, \bibinfo {author} {\bibfnamefont {S.}~\bibnamefont {Chen}}, \bibinfo
  {author} {\bibfnamefont {Y.}~\bibnamefont {Zuo}}, \bibinfo {author}
  {\bibfnamefont {J.}~\bibnamefont {Yun}}, \bibinfo {author} {\bibfnamefont
  {B.}~\bibnamefont {Cui}}, \bibinfo {author} {\bibfnamefont {K.}~\bibnamefont
  {Wu}}, \bibinfo {author} {\bibfnamefont {X.}~\bibnamefont {Guo}}, \bibinfo
  {author} {\bibfnamefont {D.}~\bibnamefont {Yang}}, \bibinfo {author}
  {\bibfnamefont {J.}~\bibnamefont {Wang}}, \ and\ \bibinfo {author}
  {\bibfnamefont {L.}~\bibnamefont {Xi}},\ }\href {\doibase
  10.1038/s41598-018-31201-2} {\bibfield  {journal} {\bibinfo  {journal} {Sci.
  Rep.}\ }\textbf {\bibinfo {volume} {8}},\ \bibinfo {pages} {12959} (\bibinfo
  {year} {2018})}\BibitemShut {NoStop}%
\bibitem [{\citenamefont {Kim}\ \emph {et~al.}(2017)\citenamefont {Kim},
  \citenamefont {Gr{\"{u}}nberg}, \citenamefont {Han},\ and\ \citenamefont
  {Cho}}]{Kim2017a}%
  \BibitemOpen
  \bibfield  {author} {\bibinfo {author} {\bibfnamefont {T.~H.}\ \bibnamefont
  {Kim}}, \bibinfo {author} {\bibfnamefont {P.}~\bibnamefont {Gr{\"{u}}nberg}},
  \bibinfo {author} {\bibfnamefont {S.~H.}\ \bibnamefont {Han}}, \ and\
  \bibinfo {author} {\bibfnamefont {B.~K.}\ \bibnamefont {Cho}},\ }\href
  {\doibase 10.1038/s41598-017-04883-3} {\bibfield  {journal} {\bibinfo
  {journal} {Sci. Rep.}\ }\textbf {\bibinfo {volume} {7}},\ \bibinfo {pages}
  {4515} (\bibinfo {year} {2017})}\BibitemShut {NoStop}%
\bibitem [{\citenamefont {Kimel}\ \emph {et~al.}(2009)\citenamefont {Kimel},
  \citenamefont {Ivanov}, \citenamefont {Pisarev}, \citenamefont {Usachev},
  \citenamefont {Kirilyuk},\ and\ \citenamefont {Rasing}}]{Kimel2009}%
  \BibitemOpen
  \bibfield  {author} {\bibinfo {author} {\bibfnamefont {A.~V.}\ \bibnamefont
  {Kimel}}, \bibinfo {author} {\bibfnamefont {B.~A.}\ \bibnamefont {Ivanov}},
  \bibinfo {author} {\bibfnamefont {R.~V.}\ \bibnamefont {Pisarev}}, \bibinfo
  {author} {\bibfnamefont {P.~A.}\ \bibnamefont {Usachev}}, \bibinfo {author}
  {\bibfnamefont {A.}~\bibnamefont {Kirilyuk}}, \ and\ \bibinfo {author}
  {\bibfnamefont {T.}~\bibnamefont {Rasing}},\ }\href
  {http://www.nature.com/articles/nphys1369} {\bibfield  {journal} {\bibinfo
  {journal} {Nat. Phys.}\ }\textbf {\bibinfo {volume} {5}},\ \bibinfo {pages}
  {727} (\bibinfo {year} {2009})}\BibitemShut {NoStop}%
\bibitem [{\citenamefont {Mondal}\ \emph {et~al.}(2017)\citenamefont {Mondal},
  \citenamefont {Berritta}, \citenamefont {Nandy},\ and\ \citenamefont
  {Oppeneer}}]{Mondal2017}%
  \BibitemOpen
  \bibfield  {author} {\bibinfo {author} {\bibfnamefont {R.}~\bibnamefont
  {Mondal}}, \bibinfo {author} {\bibfnamefont {M.}~\bibnamefont {Berritta}},
  \bibinfo {author} {\bibfnamefont {A.~K.}\ \bibnamefont {Nandy}}, \ and\
  \bibinfo {author} {\bibfnamefont {P.~M.}\ \bibnamefont {Oppeneer}},\ }\href
  {\doibase 10.1103/PhysRevB.96.024425} {\bibfield  {journal} {\bibinfo
  {journal} {Phys. Rev. B}\ }\textbf {\bibinfo {volume} {96}},\ \bibinfo
  {pages} {024425} (\bibinfo {year} {2017})}\BibitemShut {NoStop}%
\bibitem [{\citenamefont {Manchon}\ and\ \citenamefont
  {Belabbes}(2017)}]{Manchon2017}%
  \BibitemOpen
  \bibfield  {author} {\bibinfo {author} {\bibfnamefont {A.}~\bibnamefont
  {Manchon}}\ and\ \bibinfo {author} {\bibfnamefont {A.}~\bibnamefont
  {Belabbes}},\ }\href@noop {} {\bibfield  {journal} {\bibinfo  {journal}
  {Solid State Phys.}\ }\textbf {\bibinfo {volume} {68}},\ \bibinfo {pages} {1}
  (\bibinfo {year} {2017})}\BibitemShut {NoStop}%
\bibitem [{\citenamefont {Ambrosch-Draxl}\ and\ \citenamefont
  {Sofo}(2006)}]{Ambrosch-Draxl2006}%
  \BibitemOpen
  \bibfield  {author} {\bibinfo {author} {\bibfnamefont {C.}~\bibnamefont
  {Ambrosch-Draxl}}\ and\ \bibinfo {author} {\bibfnamefont {J.~O.}\
  \bibnamefont {Sofo}},\ }\href {\doibase 10.1016/J.CPC.2006.03.005} {\bibfield
   {journal} {\bibinfo  {journal} {Comp. Phys. Commun.}\ }\textbf {\bibinfo
  {volume} {175}},\ \bibinfo {pages} {1} (\bibinfo {year} {2006})}\BibitemShut
  {NoStop}%
\bibitem [{\citenamefont {Perdew}\ \emph {et~al.}(1996)\citenamefont {Perdew},
  \citenamefont {Burke},\ and\ \citenamefont {Ernzerhof}}]{Perdew1996}%
  \BibitemOpen
  \bibfield  {author} {\bibinfo {author} {\bibfnamefont {J.~P.}\ \bibnamefont
  {Perdew}}, \bibinfo {author} {\bibfnamefont {K.}~\bibnamefont {Burke}}, \
  and\ \bibinfo {author} {\bibfnamefont {M.}~\bibnamefont {Ernzerhof}},\ }\href
  {\doibase 10.1103/PhysRevLett.77.3865} {\bibfield  {journal} {\bibinfo
  {journal} {Phys. Rev. Lett.}\ }\textbf {\bibinfo {volume} {77}},\ \bibinfo
  {pages} {3865} (\bibinfo {year} {1996})}\BibitemShut {NoStop}%
\bibitem [{\citenamefont {Oppeneer}(2001)}]{Oppeneer2001}%
  \BibitemOpen
  \bibfield  {author} {\bibinfo {author} {\bibfnamefont {P.~M.}\ \bibnamefont
  {Oppeneer}},\ }in\ \href@noop {} {\emph {\bibinfo {booktitle} {Handbook of
  Magnetic Materials}}},\ Vol.~\bibinfo {volume} {13},\ \bibinfo {editor}
  {edited by\ \bibinfo {editor} {\bibfnamefont {K.~H.~J.}\ \bibnamefont
  {Buschow}}}\ (\bibinfo  {publisher} {Elsevier, Amsterdam},\ \bibinfo {year}
  {2001})\ pp.\ \bibinfo {pages} {229 -- 422}\BibitemShut {NoStop}%
\bibitem [{\citenamefont {Monkhorst}\ and\ \citenamefont
  {Pack}(1976)}]{Monkhorst1976}%
  \BibitemOpen
  \bibfield  {author} {\bibinfo {author} {\bibfnamefont {H.~J.}\ \bibnamefont
  {Monkhorst}}\ and\ \bibinfo {author} {\bibfnamefont {J.~D.}\ \bibnamefont
  {Pack}},\ }\href {\doibase 10.1103/PhysRevB.13.5188} {\bibfield  {journal}
  {\bibinfo  {journal} {Phys. Rev. B}\ }\textbf {\bibinfo {volume} {13}},\
  \bibinfo {pages} {5188} (\bibinfo {year} {1976})}\BibitemShut {NoStop}%
\end{thebibliography}
%

\end{document}